\documentclass[a4paper,11pt]{article}
\usepackage[english]{babel}
\usepackage{amsmath}
\usepackage{epsfig}
\usepackage[a4paper,bookmarksnumbered,bookmarksopen,bookmarksopenlevel=1,colorlinks=true]{hyperref}
\usepackage[dvips]{color}
\definecolor{darkred}{rgb}{0.5,0,0}
\definecolor{darkgreen}{rgb}{0,0.5,0}
\definecolor{darkblue}{rgb}{0,0,0.5}
\hypersetup{colorlinks
  ,linkcolor=darkblue
  ,filecolor=darkgreen
  ,urlcolor=darkred
  ,citecolor=darkblue}
\hypersetup{
  pdftitle={Fittino, a program for determining MSSM parameters from collider observables using an iterative method},%
  pdfauthor={Philip Bechtle, Klaus Desch, Peter Wienemann},%
}
\newcommand{\ra} {\ensuremath{\rightarrow}}

\newcommand{\ee}{\mbox{${\mathrm{e}}^+ {\mathrm{e}}^-$}}

\newcommand{\mZ}         {\mbox{$m_{\mathrm{Z}}$}}

\newcommand{\mA}         {\mbox{$m_{\mathrm{A}}$}}

\newcommand{\tanb}       {\mbox{$\tan\beta$}}

\newcommand{\lsim}{\;\raisebox{-0.9ex}{$\textstyle\stackrel{\textstyle<}
           {\sim}$}\;}

\newcommand{\placefigc}[2]{
  \begin{center}
    \epsfig{file=#1, width=#2\textwidth, clip=}
  \end{center}
}
\newcommand{\placefigs}[2]{
  \epsfig{file=#1, width=#2\textwidth, clip=}
}

\begin{document}

%------------------------------------------------------------------------------------
%----------------------------------------------
% title page
\begin{titlepage}
\begin{center}
\thispagestyle{empty}
\qquad\bigskip\bigskip\bigskip\bigskip\\
{\bf\Huge Fittino, a program\\[-1mm] for determining MSSM parameters\\[-1mm] from collider observables\\[3.5mm] using an iterative method}\bigskip\bigskip\bigskip\\
{\large P.~Bechtle$^{a,b}$, K.~Desch$^{a}$ and P.~Wienemann$^{b}$ }\bigskip\bigskip\bigskip\bigskip\\
%{\Large }\bigskip\bigskip\\
{\large $^{a}$Institut f\"ur Experimentalphysik, Universit\"at Hamburg, Luruper Chaussee 149, D-22761 Hamburg, Germany\\
$^{b}$Deutsches Elektronen-Synchrotron DESY, Notkestr. 85, D-22607 Hamburg, Germany}\bigskip\bigskip\bigskip\\
\end{center}
\begin{abstract}
  Provided that Supersymmetry (SUSY) is realized, the Large Hadron
  Collider (LHC) and the future International Linear Collider (ILC)
  may provide a wealth of precise data from SUSY processes.  An
  important task will be to extract the Lagrangian parameters. On this
  basis the goal is to uncover the underlying symmetry breaking
  mechanism from the measured observables.  In order to determine the
  SUSY parameters, the program Fittino has been developed. It uses an
  iterative fitting technique and a Simulated Annealing algorithm to
  determine the SUSY parameters directly from the observables without
  any {\it a priori} knowledge of the parameters, using all available
  loop-corrections to masses and couplings. Simulated Annealing is
  implemented as a stable and efficient method for finding the optimal
  parameter values. The theoretical predictions can be provided from
  any program with SUSY Les Houches Accord interface. As fit result, a
  set of parameters including the full error matrix and
  two-dimensional uncertainty contours are obtained. Pull
  distributions can automatically be created and allow an independent
  cross-check of the fit results and possible systematic shifts in the
  parameter determination.  A determination of the importance of the
  individual observables for the measurement of each parameter can be
  performed after the fit.  A flexible user interface is implemented,
  allowing a wide range of different types of observables and a wide
  range of parameters to be used.
% This method is used successfully in first
%  general MSSM fits for the SPS1a scenario.  The results underline the
%  importance of having both access to a large part of the
%  supersymmetric particle spectrum (as provided by the Linear Collider
%  and the Large Hadron Collider LHC together) and -- at least partly
%  -- very precise measurements of SUSY observables (as obtained at a
%  future Linear Collider).
\end{abstract}
\end{titlepage}

%------------------------------------------------------------------------------------
%------------------------------------------------------------------------------------
% table of contents
%\tableofcontents 
%\clearpage

% Aenderungen fuer Fittino Techical Paper:
% 
% - Fittino-Versions-Referenz auf 1.0.3 aendern.
% - BR-Alias hinter "BR" geloescht, nur noch hinter dem Wert wird Alias angegeben.
% - brsum eingefuehrt
% - Fuer sigma, BR, brsum muessen keine Werte mehr angegeben werden (fuer reine Alias-Defininition).
% - Input-File wird als erster Programmparameter Fittino uebergeben, Default-Setting: "fittino.in".
% - sqrt(s)-Angabe bei Cross-Sections kann auch mit Einheit erfolgen. Default: GeV.
% - Verbose mode
% - Universality kann eine ganze Liste mit Parametern folgen
% - scanParameters on/off
% - scanParameter MSelectronR  115.1 GeV  116.1 GeV  20
% - PerformFit on/off
% 
% - TTree in PullDistributions.root File
% - Fuer fitParameter muss kein Wert angegeben werden.
% - Inputparameter RandomGeneratorSeed eingebaut.
%   Falls er < 0 ist, wird ein Random-Seed aus uptime, freeswap, pid und systime genommen.
%   Andernfalls wird der angegebene Wert verwendet.
% 
% - (Vielleicht nicht erwaehnen: SimAnnUncertainty und RandomDirUncertainty eingebaut).

%------------------------------------------------------------------------------------
%------------------------------------------------------------------------------------
% text

\section{Introduction}\label{sec:intro}
Fittino~\cite{Fittinowebsite} is a program implemented in
$\mathrm{C}\!\!+\!\!+$, which extracts the parameters of the SUSY
Lagrangian from (simulated) measurements at future colliders such as
the LHC and the ILC in a global fit. It has been created in the
context of the Supersymmetry Parameter Analysis
(SPA)~\cite{SPAwebsite,SPAdraft} project. In Fittino, no \emph{a
  priori} knowledge of the parameters is assumed. In contrast to the
approach pursued in~\cite{Plehn:2004rp,SFitterwebsite}, tree-level
relations among observables and subsets of SUSY parameters are used
to obtain start values for the $\chi^2$-fit. Fits in parts of the
parameter space, using only observables directly depending on the
fitted subset of parameters, are used to refine the tree-level
parameter estimates.  Alternatively, a Simulated Annealing algorithm
can be used to globally minimise the $\chi^2$ for all parameters with
respect to all observables. Finally, a global fit and an uncertainty
analysis provide the best parameter values, their uncertainties and
their correlations.  Additionally, two-dimensional graphical contours
of the parameter uncertainties can be obtained, pull distributions and
$\chi^2$ distributions can be automatically created and an analysis of
the importance of individual observables for the determination of each
parameter can be performed.

Fittino allows to perform a simultaneous fit of any combination of 24
parameters of the low energy MSSM Lagrangian and 7 Standard Model (SM)
parameters. Alternatively high-scale mSUGRA, GMSB or AMSB parameters
can be fitted.  As observables, measurements of present or future
collider experiments can be used in the fit, such as masses,
cross-sections, branching fractions, widths, products of
cross-sections and branching fractions, ratios of branching fractions
and edges in mass spectra.  Correlations among observables can be
specified. This allows a realistic test of the precision of the
parameter determination of the low-energy MSSM theory at future
colliders. The data of measured and simulated observables and their
uncertainties (if desired including theoretical uncertainties) and
correlations is entered by the user.  The theory prediction for these
observables is calculated by a theory code, which is interfaced to
Fittino via the SUSY Les Houches Accord
(SLHA)~\cite{Skands:2003cj}. In the present implementation,
SPheno~\cite{SPHENO} is used as an example to calculate the observable
predictions from the set of parameters.

The user interface of Fittino allows the user to specify any number of
observables and parameters as given in the list above. The behaviour
of Fittino can be steered with a wide range of options. Start values for
the global fit can either be determined by Fittino, using tree-level
estimates, several sequential fits in parts of the parameter space
(subsector fit) or Simulated Annealing, or they can be specified by
the user if desired.

%Without the information from all sectors of the
%theory this fit does not converge. Therefore both the almost complete
%spectrum at LHC and the precise measurements of the lighter SUSY
%particles at the ILC is crucial. Without the very precise measurements
%of the ILC the parameter correlations would be too large to correctly
%calculate correlation matrices, even if the fit converges correctly.
%Therefore the mutual benefit of the LHC and the ILC is obvious in this
%study. 
%A model fit is performed for the benchmark parameter set
%SPS1a~\cite{Allanach:2002nj}, assuming unification in the first two
%generations. As a result of the fit, a full error matrix of the
%parameters and two-dimensional uncertainty contours of the parameters
%are obtained.

Fittino has been tested for different MSSM scenarios. The fit
strategies of Fittino, both with subsector fits and with the Simulated
Annealing algorithm, provide stable parameter reconstruction using
simulated measurements from ILC and LHC. The results of these fits are
reported in~\cite{fittino_results}.

This paper is organised as follows: First the general concepts of MSSM
parameter determination are introduced, followed by a description of
the fit program Fittino. A documentation of Fittino is given starting
from Section~\ref{app:fittino_docu}, where the syntax and the commands
of the input file are discussed. In Section~\ref{sec:quickstart}, a
short guideline for how to start a fit with Fittino is given. This is
followed by a description of the output in the remaining chapters. In
Section~\ref{sec:summary}, a summary and a short description of the
results obtained with Fittino for various MSSM
scenarios~\cite{fittino_results} are given. All information provided
in this article refers to Fittino version 1.1.1.
\section{The Fit Program Fittino}\label{sec:fittino_itself}

In the presence of experimental uncertainties which are much larger
than loop effects in a theory, parameters can be measured using
analytical tree-level relations among parameters and observables.
However, for very good experimental accuracy this method is
depreciated for a correct parameter and uncertainty determination,
since parameters tend to be systematically off their true values and
since correlations can not be fully taken into account. This plays a
role if the relative uncertainties of the measured observables $O_i$
are smaller than the largest contribution from loop effects
$$\frac{\Delta O_i^{\mathrm{meas}}}{O_i}\lsim\frac{\Delta
  O_i^{\mathrm{loop}}}{O_i},$$
as in case of the future measurements
of the ILC. 
%If the uncertainties of the measurements are much larger
%than the expected contribution from loop effects, the tree-level
%parameter determination is sufficient.  
Therefore in this case all
available loop corrections have to be taken into account, in order to
achieve the highest possible precision. This means that all
observables $O_i$ are treated as functions of all parameters $P_j$:
\begin{center}
  Observable $O_i=f(\mathrm{all \, parameters}\, P_j)$.
\end{center}
Additionally, in order to account for the uncertainties from the
limited precision of SM parameters (parametric uncertainties), the SM
parameters can be fitted simultaneously with MSSM parameters.  This
approach also allows to extract the full correlation among all
parameters. No bias is introduced due to {\it a priori} assumptions on
fixed parameters.

\subsection{The Fittino Approach}

The aim of Fittino is the unbiased determination of the parameters of
the MSSM Lagrangian ${\cal L}_{\mathrm{MSSM}}$, obeying the following
principles:
\begin{itemize}
\item No {\it a priori} knowledge of SUSY parameters is assumed (but can be
  used if desired by the user).
\item All measurements from future colliders could be used.
\item All correlations among parameters and all influences of
  loop-induced effects are taken into account.
\end{itemize}

It is impossible to determine all 105 possible parameters of ${\cal
  L}_{\mathrm{MSSM}}$ simultaneously. Therefore, assumptions on the
structure of ${\cal L}_{\mathrm{MSSM}}$ are made.  All complex phases
are set to 0, no mixing between generations is assumed and the mixing
within the first two generations is set to 0. Thus the number of free
parameters is reduced to 24 (MSSM-24).  Further assumptions can be
specified by the user.  Observables used in the fit can be
\begin{itemize}
\item Masses, limits on masses of unobserved particles
\item Widths
\item Cross-sections (momentarily in $\ee$ collisions only)
\item Branching fractions
\item Edges in mass spectra of decay products. For example in slepton
  decays $\tilde{\mu}^+\tilde{\mu}^-\ra\mu^+\chi^0_1\mu^-\chi^0_1$,
  the lepton energy spectrum has a box-like shape. Its edges are
  correlated to the masses of the $\tilde{\mu}$ and the $\chi^0_1$. By
  using the edge positions instead of the reconstructed masses,
  correlations among the observables can be reduced or omitted (see e.g.
  \cite{Martyn:1999xc}).
\item Products of cross-sections and branching fractions. For most
  processes, it is hard to measure total cross-sections. Therefore an
  observable can be formed as a product of a cross-section and several
  branching fractions.
\item Ratios of branching fractions. If the total width of a decay can
  not be measured, the individual measured widths can be all
  normalised to one width and thus form ratios of branching fractions.
\end{itemize}
Correlations among observables and both experimental and theoretical
errors can be specified. Theoretical uncertainties can be important,
if they are in the order of magnitude of the experimental
uncertainties.  Both SM and MSSM observables can be used in the fit.
Parametric uncertainties of SUSY observables can be taken into account
by fitting the relevant SM parameters simultaneously with the MSSM-24,
the mSUGRA, the GMSB or the AMSB parameters.  For the interface
between Fittino and the code providing the theoretical predictions,
the SUSY Les~Houches Accord~\cite{Skands:2003cj} (SLHA), is used. The
SLHA is a format for a text-file based interface between spectrum
calculators, event generators and other programs in the context of the
MSSM. Any theoretical code compliant with SLHA can be easily
interfaced with Fittino. In the current implementation, the prediction
of the MSSM observables for a given set of parameters is obtained from
SPheno~\cite{SPHENO}. A Simulated Annealing algorithm based
on~\cite{SimAnn,SimAnn2} and MINUIT~\cite{James:1975dr} is used for
the fitting process.
%
%In the following, the principles of Fittino are outlined in more
%detail, followed by an example for a fit based on
%SPS1a~\cite{Allanach:2002nj}.
%
%
%In the following, the program Fittino is described, which is able to
%determine the 24 parameters of the SUSY Lagrangian without {\it a priori}
%assumptions on the parameters. In order to account for parametric
%uncertainties, SM parameters can be fitted simultaneously. First the
%fit procedure and the fit method is explained, followed by an
%introduction into the steering of Fittino using ASCII input files.

\subsection{The Iterative Fit Procedure}

The full MSSM-24 parameter space in Fittino, consisting of MSSM
parameters plus SM parameters, cannot be scanned completely, neither
in a fit nor in a grid approach.  Therefore, in order to find the true
parameter set $\vec{P}$ in the presence of $n_{\mathrm{obs}}$ measured
observables $O_i^{\mathrm{meas}}$ in a fit by minimising a $\chi^2$
function
($\chi^2=\sum_{i=1}^{n_{\mathrm{obs}}}(O_i^{\mathrm{meas}}-O_i^{\mathrm{theo}}(\vec{P}))^2/\sigma_i^2$),
it is essential to begin with reasonable start values, allowing for a
smooth transition to the true minimum.  The three different tasks in
the parameter determination are:
\begin{enumerate}
\item \emph{Tree-level estimate}\\
  Using a small number of observables, the initial values for the
  parameters are determined on tree-level. 
\item \emph{Finding the central values of the parameters}\\
  Starting from the initial values of the parameters, the
  configuration of parameter values with the smallest total $\chi^2$,
  i.e.{} the best agreement between theory predictions and measurements,
  has to be found. 
\item \emph{Uncertainty determination}\\
  Once the optimal parameter values are found, the evaluation of their
  correlations and uncertainties is performed.
\end{enumerate}

%\begin{figure}[t]
%\begin{center}
%\epsfig{file=figures/schema.eps, width=0.8\textwidth,clip=}
%\end{center}
%\caption[The iterative fit procedure of Fittino]{\sl The 
%  iterative fit procedure of Fittino. From the observables $O_i$ and
%  their uncertainties $\Delta O_i$, the start values for parameters
%  $P_j$ of the fit are calculated in a bias-free way using tree-level
%  relations. After the fit, the fitted parameters $P_j$ and their
%  uncertainties $\Delta P_j$ are obtained, including their full
%  correlation matrix.}\label{fig:iterative}
%\end{figure}
%
\begin{figure}[p]
\begin{center}
\epsfig{file=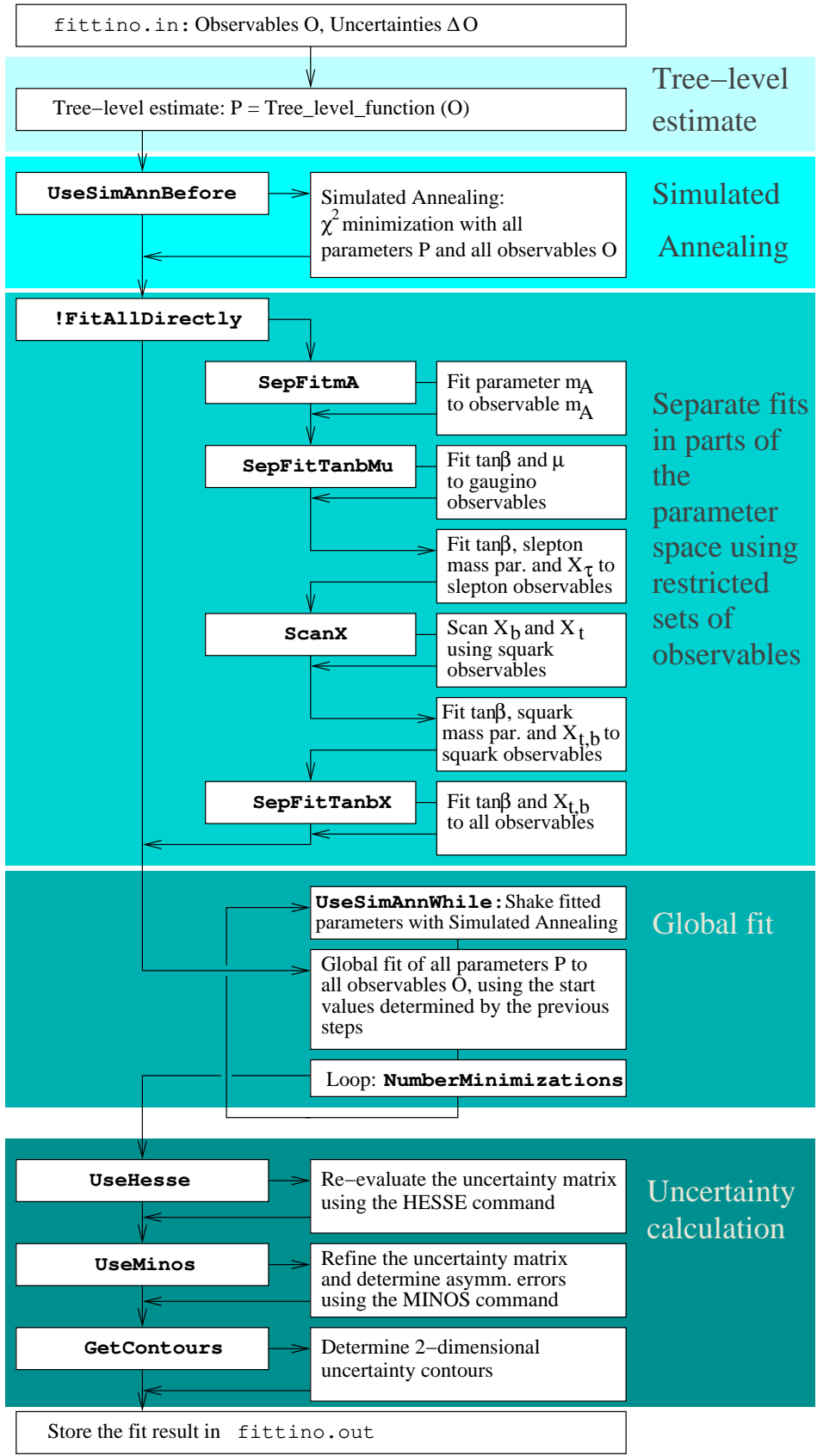, width=0.8\textwidth,clip=}\vspace{-5mm}
\end{center}
\caption[Flow Chart of the individual fit steps in Fittino]{\sl Flow chart of the 
  iterative fit procedure of Fittino. Output and input files are shown
  in light {\tt courier}, boolean operations in bold
  {\bf\tt\bf\tt courier} and explanations are shown in {\rm times}.
  }\label{fig:flowchart}
\end{figure}

The iterative procedure of Fittino to determine the start values for
the fit, the optimal parameter values and the parameter uncertainties
is displayed in Fig.~\ref{fig:flowchart}. 
%In more detail, the
%functionality of Fittino and the order of the different operations is
%shown in Fig.~\ref{fig:flowchart}. 
In a first step, the SUSY parameters are estimated using tree-level
relations. Then several subsequent subsector fits in parts of the
parameter space can be used to improve the tree level estimates before
the global fits. This is called the subsector fit method.
Alternatively, a Simulated Annealing~\cite{SimAnn,SimAnn2} algorithm
can be used to find the minimum. In a third step a global fit of all
parameters to all observables is used to fine-tune the optimal
parameter values and determine the parameter uncertainties and
correlations. 

%----------------------------------------------------------------------
\subsubsection{Tree level estimates for the parameters}\label{sec:tree_level_est}

In order to get the start values for both the subsector fit method and the
Simulated Annealing, tree-level relations of the form 
$$\mathrm{Observable}=f(\mathrm{Parameters})$$
are inverted to the form
$$\mathrm{Parameter}=f'(\mathrm{Observables})$$
and used to estimate
the parameters. This is done in the gaugino, the slepton and the
squark sectors.
\begin{enumerate}
\item $\mu,m_A,\tan\beta,M_1,M_2,M_3$ are determined from the gaugino
  and Higgs sector using formulae
  from~\cite{Zerwas:2002as,Desch:2003vw}.  In order to extract these
  parameters, information from chargino cross-sections is needed,
  which enters in form of the chargino mixing angles $\cos2\phi_L$ and
  $\cos2\phi_R$.  These pseudo-observables are just used for the
  determination of the start values, no use is made of them for the
  fit. They can be approximately determined from chargino
  cross-sections at the ILC at different polarisations. If
  kinematically accessible, $\mA_{\mathrm{pole}}$ can be directly
  measured. The start value of the parameter $\mA_{\mathrm{run}}$ is
  set to $\mA_{\mathrm{pole}}$ in this case.  For the other
  parameters, the following individual calculations are performed. For
  the full form of the relations, see e.~g.~\cite{SPHENO}. Based on
  the diagonalisation of the chargino mixing matrix the mixing angles
  $\phi_L$ and $\phi_R$ can then be used to derive the tree-level
  estimates of $\mu$, $\tanb$ and $M_2$:
  \begin{eqnarray}
    |\mu|      & = & m_{W}\left(\Sigma+\Delta(\cos2\phi_L+\cos2\phi_R)\right)^{\frac{1}{2}}\label{eqn:mu_det}\\
    \tanb      & = & \left(\frac{1+\Delta(\cos2\phi_R-\cos2\phi_L)}{1-\Delta(\cos2\phi_R-\cos2\phi_L)}\right)^{\frac{1}{2}} \label{eqn:tanb_det}\\
    M_2        & = & m_{W}\left(\Sigma-\Delta(\cos2\phi_L+\cos2\phi_R)\right)^{\frac{1}{2}}\\
    \text{sign}(\mu) & = & -\frac{\Delta^2-(\mu^2-M_2^2)^2-4m_W^2(\mu^2+M_2^2)-4m_W^2\cos^22\beta}{8m_W^2M_2|\mu|\sin2\beta}\\
    M_3        & = & m_{\tilde{\mathrm{g}}}
  \end{eqnarray}
  using 
  \begin{eqnarray}
    \Sigma&=&\frac{m^2_{\chi^{\pm}_2}+m^2_{\chi^{\pm}_1}}{2m^2_{W}}-1\\
    \Delta&=&\frac{m^2_{\chi^{\pm}_2}-m^2_{\chi^{\pm}_1}}{4m^2_{W}}
  \end{eqnarray}  
%  \begin{figure}[t]
%    \begin{center}
%      \epsfig{file=figures/c2lrmin.eps, width=0.5\textwidth,clip=}
%    \end{center}
%    \caption[The approximate initial determination of the chargino mixing angles]{\sl The 
%      approximate initial determination of the chargino mixing angles
%      from LR and RL chargino cross-sections. From the measurements of
%      three chargino cross-sections at different beam polarisation,
%      the chargino mixing angles can be initially
%      determined~\cite{Zerwas:2002as}.}\label{fig:cos2phi}
%  \end{figure}
  After these parameters have been determined at tree-level, a
  tree-level estimate of $M_1$ can be calculated from the neutralino
  mass matrix  eigenvalues. They depend on
  $\mu, \tanb, M_2$ and $M_1$. The first three of these parameters
  have been determined already in the first step. Thus the neutralino
  system can be used to finally determine $M_1$. In fact, the
  neutralino system is the only sector where $M_1$ can be measured
  directly. The characteristic equation of the neutralino mass matrix
  then is used to determine $M_1$~\cite{Desch:2003vw}.
\item $X_{\mathrm{t}},X_{\mathrm{b}},M_Q,M_U,M_D$ are determined from
  the squark sector masses, using formulae from~\cite{Porod:1998kk}.
  The trilinear coupling parameters $A$ are set to
  zero for the determination of the squark mass parameters.
  $X_{\mathrm{t}}=A_{\mathrm{t}}-\mu/\tanb$ and
  $X_{\mathrm{b}}=A_{\mathrm{b}}-\mu\tanb$ are chosen as fit
  parameters instead of $A$ because their correlation with $\tanb$ is
  reduced.  After $\tanb$ and $\mu$ have been determined in the
  previous step, the following tree-level relations can be used. As an
  example, only the third generation is shown explicitly.
  \begin{eqnarray}
    M_{\tilde{\mathrm{t}}_L} & = & -\mZ^2\cos2\beta (\frac{1}{2}-\frac{2}{3}\sin^2\theta_W)-
    m_{\mathrm{t}}^2+\frac{1}{2}(m^2_{\tilde{\mathrm{t}}_1}+m^2_{\tilde{\mathrm{t}}_2}) \\
                     & = & \mZ^2\cos2\beta (\frac{1}{2}-\frac{1}{3}\sin^2\theta_W)-
    m_{\mathrm{b}}^2+\frac{1}{2}(m^2_{\tilde{\mathrm{b}}_1}+m^2_{\tilde{\mathrm{b}}_2}) \\
    M_{\tilde{\mathrm{t}}_R} & = & -\mZ^2\cos2\beta \frac{2}{3}\sin^2\theta_W-
    m_{\mathrm{t}}^2+\frac{1}{2}(m^2_{\tilde{\mathrm{t}}_1}+m^2_{\tilde{\mathrm{t}}_2}) \\
    M_{\tilde{\mathrm{b}}_R} & = & \mZ^2\cos2\beta \frac{1}{3}\sin^2\theta_W-
    m_{\mathrm{b}}^2+\frac{1}{2}(m^2_{\tilde{\mathrm{b}}_1}+m^2_{\tilde{\mathrm{b}}_2}) \\
    X_{\mathrm{t}}          & = & -\mu/\tanb \\
    X_{\mathrm{b}}            & = & -\mu\tanb  
  \end{eqnarray}
  The parameters of the other generations can be obtained analogously.
  The initial determination of $X_{\mathrm{t}}$ and
  $X_{\mathrm{b}}$ is only very rough and therefore refined in an
  additional step, which is explained below.
% Formulae!
\item $X_{\tau},M_L,M_E$ are determined from the slepton sector
  masses, using formulae from~\cite{Porod:1998kk}.  Their calculation
  is analogous to the calculation of the squark parameters in the
  previous item.
%  Here $X_{\tau}=A_{\tau}-\mu\tanb$
%  from
%  \begin{eqnarray}
%    M^2_{\tilde\nu} & = & M^2_{\tau_L} + \frac{1}{2} m^2_Z\cos2\beta \label{eqn:slepton_mixing_a}\\
%    M^2_{\tilde \tau}  & = & \matrixtwo
%    {M^2_{\tau_L}+ m^2_\tau- m^2_Z(\frac{1}{2}- s^2_W)\cos2\beta}
%    {m_\tau(A_\tau-\mu\tan\beta)}
%    {m_\tau(A_\tau-\mu\tan\beta)}
%    {M^2_{\tau_R}+ m^2_\tau- m^2_Z s^2_W\cos2\beta}
%    \label{eqn:slepton_mixing}
%  \end{eqnarray}
%  is set to zero in the tree-level estimate, too. The tree-level
%  formulae derived from (\ref{eqn:slepton_mixing_a}) and
%  (\ref{eqn:slepton_mixing}) read for the third generation:
%  \begin{eqnarray}
%    M_{\tilde{\mathrm{t}}_L} & = & \mZ^2\cos2\beta (\frac{1}{2}-\sin^2\theta_W)-
%    m_{\tau}^2+\frac{1}{2}(m^2_{\tilde{\tau}_1}+m^2_{\tilde{\tau}_2}) \\
%    M_{\tilde{\mathrm{t}}_R} & = & -\mZ^2\cos2\beta \sin^2\theta_W-
%    m_{\tau}^2+\frac{1}{2}(m^2_{\tilde{\tau}_1}+m^2_{\tilde{\tau}_2}) \\  
%    X_{\tau}         & = & -\mu\tanb
%  \end{eqnarray}
  The determination of $X_{\tau}$ is refined later as explained below.
% Formulae!
\end{enumerate}
In order to estimate the uncertainty of this determination of the
parameters from tree-level formulae, the calculation is repeated
10\,000 times with observables randomly smeared within their
uncertainties according to a Gaussian probability distribution. The
starting value of the fit is the mean of the distribution of each
parameter. From the variance of the calculated parameter distribution
the initial uncertainty is estimated.

%----------------------------------------------------------------------
\subsubsection{Finding the optimal parameter values}\label{sec:parameter_improvements}

A global $\chi^2$ fit of a large number of MSSM parameters with
starting values derived from tree-level relations will most likely
fail to find the global $\chi^2$ minimum, since the parameter space
contains many local minima. Therefore the tree-level estimates have to
be improved before the global fit. This can be done by using knowledge
about the dependencies of sets of individual parameters on sets of
individual observables. This is done in the fast subsector fit method.
However, also in this case some local minima can be too deep to allow
MINUIT to escape from them. Therefore the slower but very stable
Simulated Annealing algorithm has been implemented alternatively.

\paragraph{Subsector Fit Method}
Since the possibility of mixing in the third generation has been
neglected in the calculation of the tree-level estimates, the
parameters $X_{\mathrm{t}},X_{\mathrm{b}},X_{\tau}$ are only
roughly initialised. A global fit with these starting values would
most likely not converge.  Therefore next the estimates from the
slepton sector are improved by fitting only the slepton parameters
$X_{\tau},M_L,M_E$ to the observables from the slepton sector, i.~e.
slepton masses, widths and cross-sections. Observables not directly
related to the slepton sector can degrade the fit result, since
parameters of other sectors are likely to be still wrong. In such a
case a parameter of the slepton sector will be pulled into a wrong
direction, in order to compensate for the wrong parameters of other
sectors. All parameters not from the slepton sector are fixed to their
estimated tree-level values. In this fit with reduced number of
dimensions MINUIT can handle the correlations among the parameter
better than in a global fit with all parameters free.

Then the third generation squark parameters are improved by only
fitting $X_{\mathrm{t}},X_{\mathrm{b}},M_Q,M_U,M_D$ to the
observables of the squark sector, masses, widths and cross-sections.
All other parameters are fixed to their previous values.

After this step the correlations among $\tan\beta$ and the third
generation slepton and squark parameters are still not optimally
modelled.  Therefore another intermediate step is introduced, where
$\tan\beta,X_{\mathrm{t}},X_{\mathrm{b}},X_{\tau}$ and
$M_{\tilde{\mathrm{t}}_{L,R}}$ are fitted to all observables and all other parameters are
fixed to their present values.

The subsector fit method does not yield a parameter set which
corresponds to the global minimum of the $\chi^2$. But starting from
the result of the subsector fit method, the subsequent global fit (see
Section~\ref{sec:parameter_uncertainties}) is able to converge to the
global minimum.

\paragraph{Simulated Annealing}
In the method of Simulated Annealing, the $\chi^2$ surface in the MSSM
parameter space is treated as a potential. A temperature $t$ is
defined, which according to a Boltzmann-distribution sets the
probability of a movement in the parameter space which corresponds to
an upward movement in the potential, i.e.{} which results in a larger
$\chi^2$ than the parameter setting in the step before. In the process
of Simulated Annealing, the temperature is reduced step by step.  This
ensures that large movements over high potential barriers are possible
in the early stages of the annealing process, in order to escape from
local minima. Later, the temperature is reduced and the parameter
setting is thus forced into areas with low $\chi^2$. The individual
steps of the algorithm work as follows:

\begin{figure}[t]
  \begin{center}
    \epsfig{file=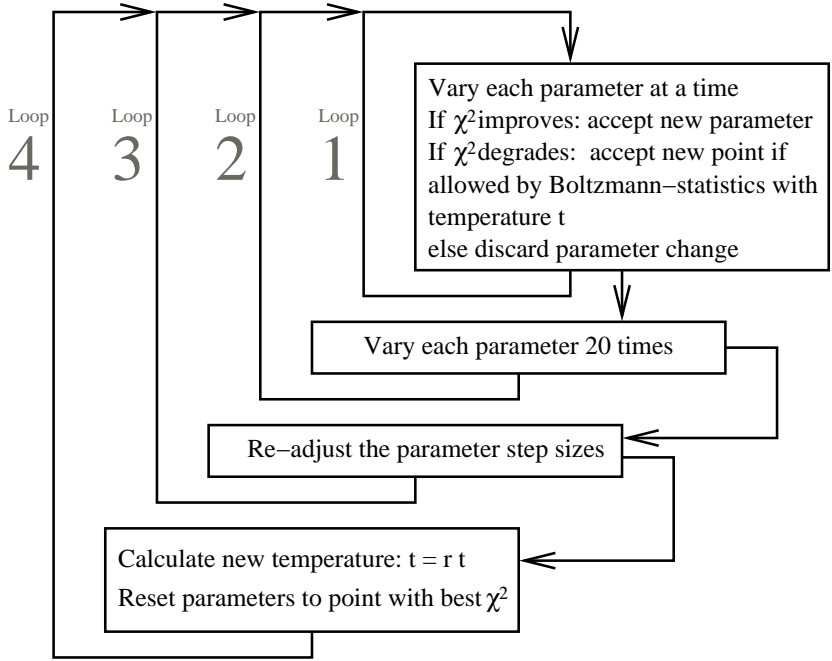, width=0.7\textwidth}
  \end{center}
  \caption[]{\sl Structure of the Simulated Annealing algorithm. 
    The four different loops are indicated by the grey numbers.}\label{fig:simannn_principle}
\end{figure}

First, the initial temperature $t_0$ is determined by varying each
parameter 10 times randomly within its estimated uncertainty. The
estimated uncertainty is determined from the tree-level estimates. Then
half of the standard deviation
$$\sigma_{\chi^2}=\sqrt{\langle(\chi^2)^2\rangle - \langle\chi^2\rangle^2}$$ 
of the $\chi^2$ values obtained is taken as
initial temperature. This choice is large enough to allow the
algorithm to scan large parts of the parameter space and to escape
from local minima. 

The core of the Simulated Annealing algorithm is schematically shown
in Fig.~\ref{fig:simannn_principle}. In the innermost loop (loop No.~1
in Fig.~\ref{fig:simannn_principle}), each parameter $P_i$ is varied
at a time by a random variation within the estimated uncertainties of
each parameter $\Delta P_i$. These uncertainties are initially taken
from the tree-level estimates and later refined for each parameter
individually, as described below. After the parameter variation, the
$\chi^2$ at the new parameter setting is evaluated. If the new value
$\chi^2_i$ lies below $\chi^2_{i-1}$, where $\chi^2_{i-1}$ is the
$\chi^2$ value of the last accepted parameter set, the new parameter
value is accepted and the next parameter is varied. If the new value
$\chi^2_i$ is the best $\chi^2$ value achieved so far, $\chi^2_i$ and
the accompanying parameter set are stored. If $\chi^2_i>\chi^2_{i-1}$,
then the new point is accepted on the basis of a Boltzmann
distribution: If a random number $p\,\epsilon\, [0,1]$ fulfils
$$p<e^{-\frac{\chi^2_i-\chi^2_{i-1}}{t}},$$
then the new parameter
value is accepted despite yielding a worse fit than the previous
parameter set. This ensures that the algorithm is able to escape from
local minima.

In loop~2 the variation of each parameter at a time is repeated
20~times. On each variation the number of accepted
($n_{\mathrm{acc}}$) and rejected
($n_{\mathrm{tot}}-n_{\mathrm{acc}}$) parameter choices is counted for
each parameter. After loop~2, the estimated parameter uncertainty
$\Delta P_i$ is iteratively re-adjusted such that
$n_{\mathrm{acc}}/n_{\mathrm{tot}}$ lies in the range of 0.4 to 0.6.
This is achieved by the transformation 
\begin{eqnarray}
\Delta P_i>0.6 & : & \Delta P^{\mathrm{new}}_i = \Delta P^{\mathrm{old}}_i\left(2\frac{\frac{n_{\mathrm{acc}}}{n_{\mathrm{tot}}}-0.6}{0.4}+1\right),\nonumber\\
\Delta P_i<0.4 & : & \Delta P^{\mathrm{new}}_i = \frac{\Delta P^{\mathrm{old}}_i}{\left(2\frac{0.4-\frac{n_{\mathrm{acc}}}{n_{\mathrm{tot}}}}{0.4}+1\right)}.\nonumber
\end{eqnarray}
In loop~3 this re-adjustment of the parameter uncertainties is
repeated $3n$ times for $n$ parameters, but at least 60 times in case
$n<20$. The re-adjustment of the parameters variation ensures that for
each temperature $t$ the random parameter variation tests a parameter
range with values of $\chi^2_1-\chi^2_{i-1}\approx t$.

After loop~3 is completed, the temperature is reduced by the
multiplication with a factor $r$, which is constant during the
algorithm. At the same time, the parameter values for the next
iteration are set to the parameter set with the lowest $\chi^2$ found
so far. The whole procedure is repeated until one of the two exit
criteria are met. Either, the allowed number of calls to the theory
code is reached, or the variation of the $\chi^2$ value within loop~3
is smaller than the variation within the last four lowest $\chi^2$
values found, and the variation of all values is smaller than 0.0001.
In each case the Simulated Annealing algorithm exits with the
parameter set which yields the best $\chi^2$ found so far.

\begin{figure}[t]
\begin{center}
% \begin{minipage}{0.49\textwidth}
%   \begin{center}
%     \epsfig{file=figures/SPS1a_tanb_chisq_4_manmade.eps, width=\textwidth}
% %    \epsfig{file=figures/SPS1a_tanb_chisq_manmade_3_klaus.eps, width=\textwidth}
% %    \includegraphics[width=\textwidth,bbllx=0,bblly=0,bburx=200,bbury=200]{figures/SPS1a_tanb_chisq_manmade_3.pdf}
% %    \epsfig{file=figures/SPS1a_tanb_chisq_manmade_3_test.eps,      width=\textwidth}
% %    \epsfig{file=figures/tanb_chisq_manmade.eps,      width=\textwidth}
% %    \epsfig{file=figures/SPS1a_tanb_chisq_manmade_3.eps, width=\textwidth}
%     (a)
%   \end{center}
% \end{minipage}
% \begin{minipage}{0.49\textwidth}
%   \begin{center}
%     \epsfig{file=figures/SPS1a_mu_n_all_manmade.eps, width=\textwidth}    
%     (b)
%   \end{center}
% \end{minipage}\\
% \begin{minipage}{0.49\textwidth}
%   \begin{center}
% \epsfig{file=figures/SPS1a_tanb_t_all_manmade.eps, width=\textwidth}
%     (c)
%   \end{center}
% \end{minipage}
% \begin{minipage}{0.49\textwidth}
%   \begin{center}
% \epsfig{file=figures/SPS1a_tanb_n_all_manmade.eps, width=\textwidth}
%     (d)
%   \end{center}
% \end{minipage}
\begin{minipage}{0.49\textwidth}
  \begin{center}
    \epsfig{file=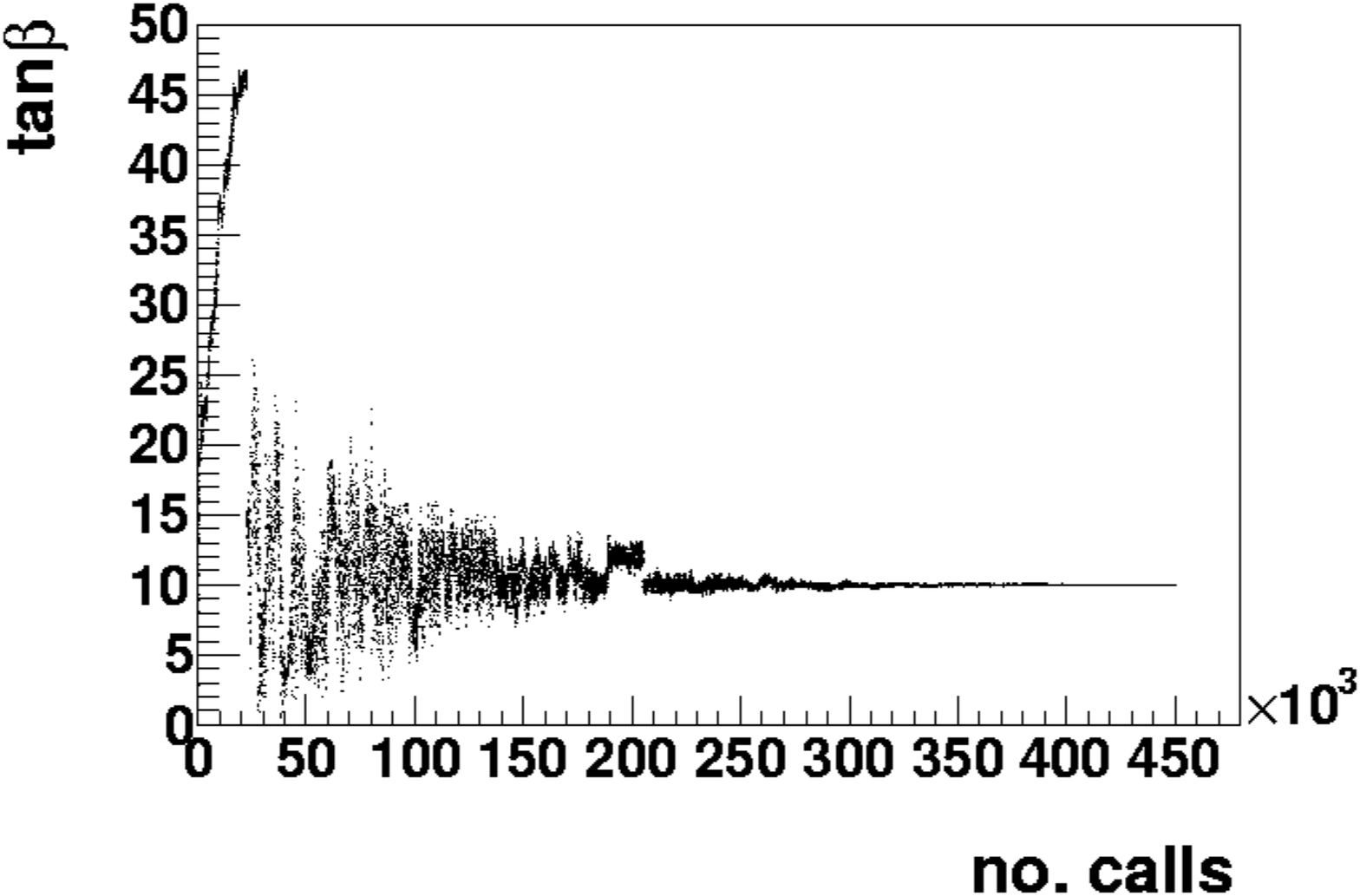, width=\textwidth}
    (a)
  \end{center}
\end{minipage}
\begin{minipage}{0.49\textwidth}
  \begin{center}
    \epsfig{file=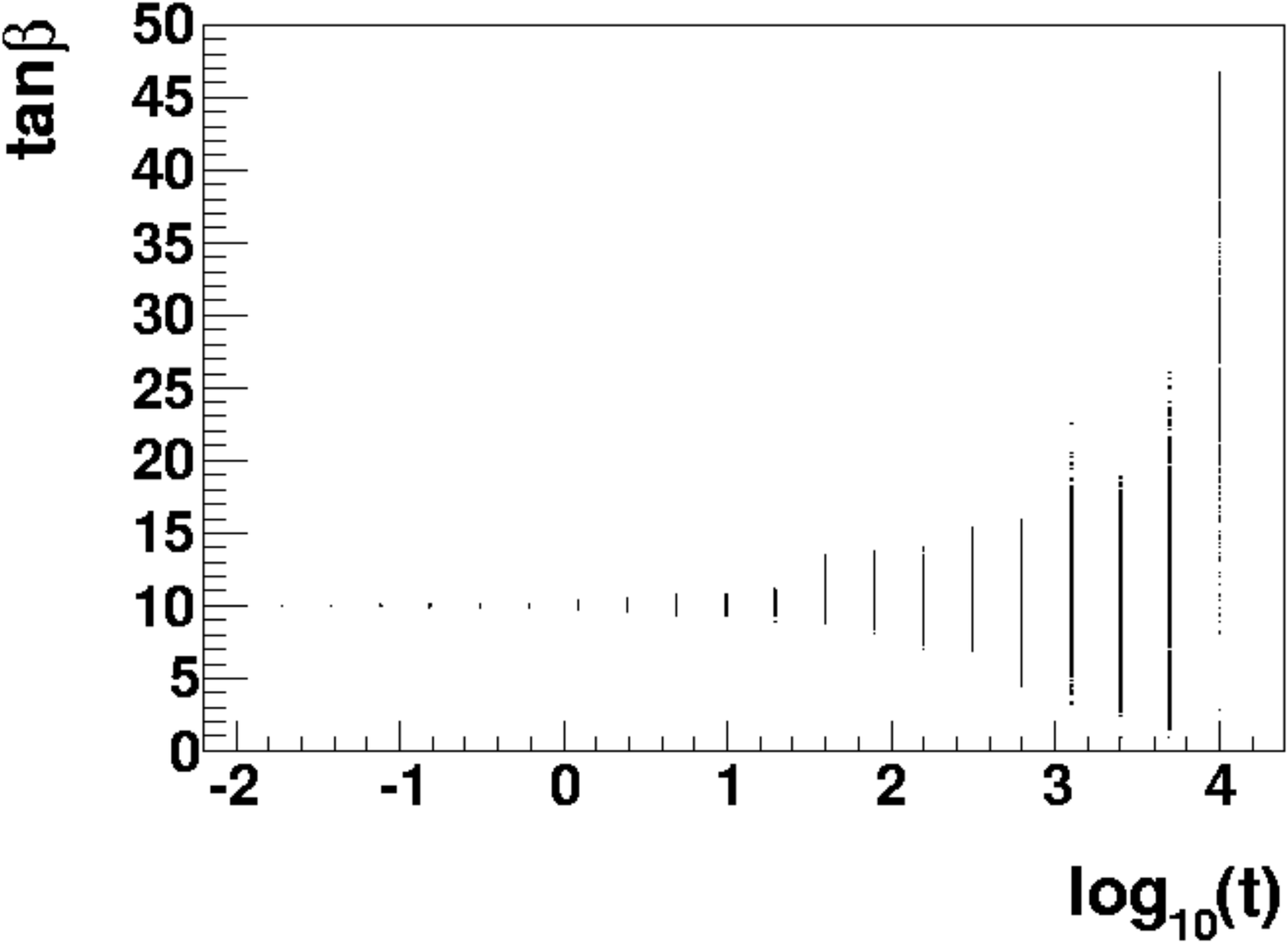, width=\textwidth}    
    (b)
  \end{center}
\end{minipage}\\
\begin{minipage}{0.49\textwidth}
  \begin{center}
\epsfig{file=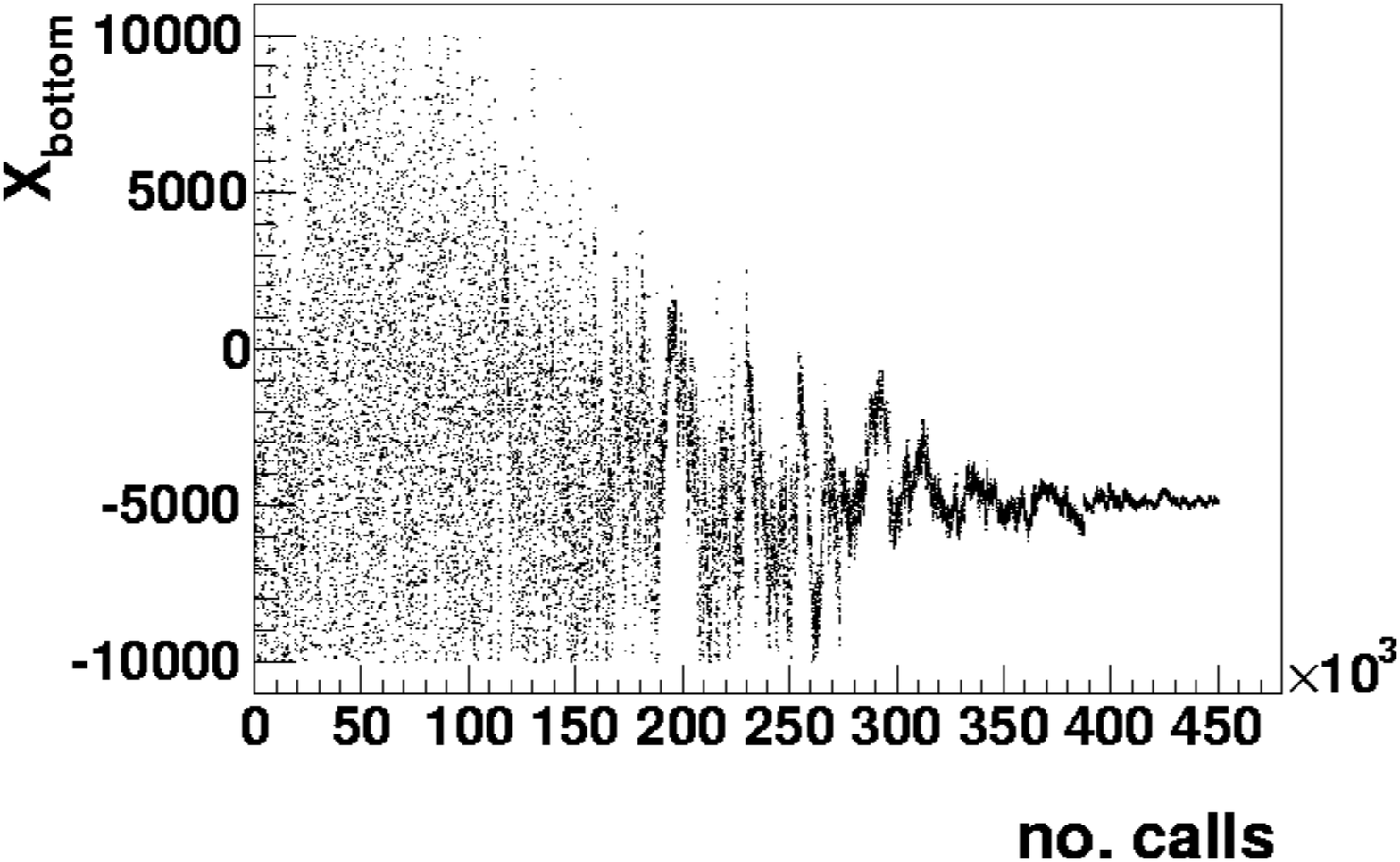, width=\textwidth}
    (c)
  \end{center}
\end{minipage}
\begin{minipage}{0.49\textwidth}
  \begin{center}
\epsfig{file=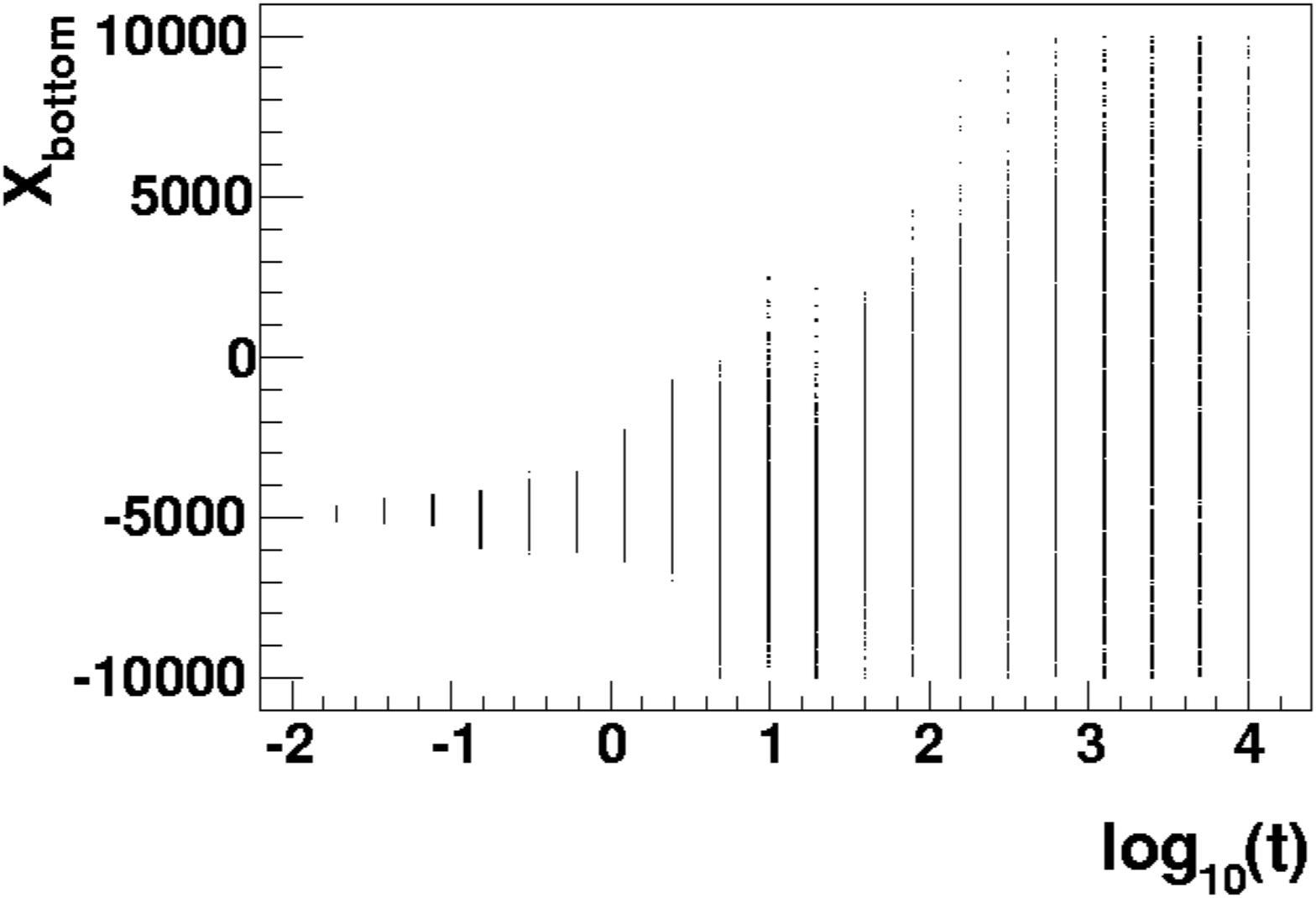, width=\textwidth}
    (d)
  \end{center}
\end{minipage}
\end{center}
\caption[]{\sl The parameter determination using Simulated Annealing 
  in a fit with 19 parameters in the SPS1a'~\cite{SPAdraft} scenario.
  In (a), the tested parameter values of $\tanb$ are shown versus the
  number of calls to SPheno. In (b) the same evolution is shown
  against $\log_{10}(t)$, where $t$ is the temperature of the
  Simulated Annealing. In (c) and (d), the same projections are shown
  for $X_{\mathrm{bottom}}$, which in contrast to $\tanb$ is only weakly
  constrained by the measurements.}\label{fig:simannn}
\end{figure}

The Simulated Annealing algorithm has been tested on fits with 20 free
parameters in the scenarios SPS1a'~\cite{SPAdraft} and
SPS7~\cite{Allanach:2002nj}. In each case, the convergence to the true
minimum of the $\chi^2$ is good and very stable. However, while being
very robust, this algorithm is considerably slower than the subsector
fit method. In Fig.~\ref{fig:simannn}, some distributions illustrating
the behaviour of the Simulated Annealing in the SPS1a' scenario are
shown.  In Fig.~\ref{fig:simannn}~(a), the variation of $\tanb$ with
the number of calls to SPheno is shown. In the beginning, a large
range of $\tanb$ is scanned. For lower $t$ at a larger number of
calls, the parameter variation $\Delta\tanb$ is reduced along with $t$
and convergence is achieved. Thus Simulated Annealing is very
effectively coarsely scanning a large parameter space at high $t$,
while it is also finely testing the area close to the $\chi^2$
minimum.  The same evolution is shown in (b) against the temperature
$t$. In (c) and (d), the same evolution is shown for the bottom-squark
mixing parameter $X_{\mathrm{bottom}}$, which is much weaker
constrained that $\tanb$ and therefore takes longer to converge. 

Simulated Annealing is a robust and reliable algorithm, which is
better capable of escaping from local minima than the subsector fit
method.  Also it is much less dependent on the starting values
obtained from the tree-level estimate.  However, it is more computing
time consuming, since a high initial temperature is necessary to
overcome the local minima near the tree-level estimates of the
parameters and find the true minimum.

%----------------------------------------------------------------------
\subsubsection{Determining the parameter uncertainties and correlations}\label{sec:parameter_uncertainties}

After the search for the optimal parameters, all parameters are
released and a global fit is done, using the method MINIMIZE in
MINUIT. In case the subsector fit method has been used for the
improvement of the tree-level estimates of the parameters, this step
is also necessary to find the optimal parameters, which are generally
not yet perfectly determined after the subsector fits. If Simulated
Annealing with a sufficient number of iterations is used, then the
optimal parameters after Simulated Annealing are generally very close
to the true parameter values.

If the global fit converges, a MINOS error analysis is performed,
yielding symmetrical and asymmetrical uncertainties, the full
correlation matrix and 2D fit contours. 

\section{The Fittino Steering File}\label{app:fittino_docu}

Fittino is controlled using an ASCII input file whose name is passed
to Fittino as parameter. If none is given, \texttt{fittino.in} is
assumed. An example for this file is given in
Section~\ref{app:fittino_input_file}. The user can specify the
observables, their values and uncertainties, their correlations, the
fitted and fixed parameters and universalities among generations.
Flags can be used to control the general behaviour of Fittino.  A list
of the available commands is given in
Tables~\ref{tab:fittino_commands} and \ref{tab:fittino_flags}. In the
following, the syntax of the commands is explained.

\begin{table}[t]
  \caption[Commands in the Fittino steering file]{\sl Commands in the Fittino steering file. All commands given in this table set the input observables and the parameters used by Fittino.
    The full form of the commands is given in the text. The correlations must be specifyed 
    after all observables. The steering flags can be found in Tab.~\ref{tab:fittino_flags}.}\label{tab:fittino_commands}
  \begin{center}
    \begin{tabular}{|l|l|}
      \hline
      Command & Explanation \\
      \hline\hline
      \multicolumn{2}{|c|}{Keys before any line}\\
      \hline\hline
      \verb|#|                     & Comment line\\
      \verb|nofit|                 & The observable after \verb|nofit| is not used in the fit\\      
      \hline\hline
      \multicolumn{2}{|c|}{Observables}\\
      \hline\hline
      \verb|mass<name>| & Mass of the particle \verb|<name>|, see Tab.~\ref{tab:fittino_particles} \\
      \verb|edge|       & Position of an edge in a mass spectrum \\
      \verb|sigma|      & cross section of a process \\
      \verb|BR|         & Branching fraction \\ 
      \verb|width|         & Total width of a particle \\
      \verb|limit|         & Limits on particle masses \\
      \verb|LEObs|      & Low energy observable \\
      \verb|sin2thetaW| & The value of $\sin^2\theta_W$\\
      \verb|cos2phiL|   & The value of $\cos2\phi_L$\\
      \verb|cos2phiR|   & The value of $\cos2\phi_R$\\
      \verb|xsbr|       & Product of cross-sections and branching fractions \\  % New!
      \verb|brratio|    & Ratio of two branching fractions \\  % New!
      \verb|brsum|      & Sum of branching fractions \\
      \hline\hline
      \multicolumn{2}{|c|}{Correlations}\\
      \hline\hline
      \verb|correlationCoefficient| & Correlation coefficient of two observables \\
      \hline\hline
      \multicolumn{2}{|c|}{Parameters}\\
      \hline\hline
      \verb|fitParameter| & Name and eventually value of a fitted parameter\\
      \verb|fixParameter| & Name and value of a fixed parameter\\
      \verb|universality| & Specifies which parameters are unified \\
      \hline
    \end{tabular}
  \end{center}
\end{table}

\begin{table}[t]
  \caption[Flags in the Fittino steering file]{\sl Flags in the Fittino steering file. The commands given here determine the fit strategies and outputs of Fittino.
    The full form of the commands is given in the text. The correlations must be specified 
    after all observables. The other commands used for the Fittino input can be found in Tab.~\ref{tab:fittino_commands}.}\label{tab:fittino_flags}
  \begin{center}
    \begin{tabular}{|l|l|}
      \hline
      \multicolumn{2}{|c|}{Flags}\\
      \hline\hline
      \verb|FitModel|              & Model type, such as MSSM, mSUGRA, etc.\\
      \verb|LoopCorrections|       & Use full loop corrections as provided by SUSY calculator \\
      \verb|ISR|                   & Switch on ISR \\
      \verb|UseGivenStartValues|   & Start from parameter values in \verb|fitParameter|\\
      \verb|FitAllDirectly|        & Fit all parameters at once \\
      \verb|CalcPullDist|          & Calculate pull distributions \\
      \verb|CalcIndChisqContr|     & Calculate individual $\Delta\chi^2$ contributions \\
      \verb|BoundsOnX|             & Set bounds on $X_{\tau}$, $X_{\mathrm{t}}$ and $X_{\mathrm{b}}$ \\       
      \verb|ScanX|                 & Scan $X_{\mathrm{t}}$ and $X_{\mathrm{b}}$ before fitting \\ 
      \verb|SepFitTanbX|           & Perform separate fit of $\tanb$, $M_{\tilde{\mathrm{t}}_R}$, $M_{\tilde{\mathrm{t}}_L}$, $X_{\mathrm{t}}$ and $X_{\mathrm{b}}$ \\
      \verb|SepFitTanbMu|           & Perform separate fit of $\tanb$ and $\mu$  \\ %New!
      \verb|SepFitmA|              & Perform separate fit of $\mA_{\mathrm{run}}$ to $\mA_{\mathrm{pole}}$  \\ %New!
      \verb|Calculator|            & Specify calculator for theory predictions\\
      \verb|UseMinos|              & Use MINOS error calculation\\  
      \verb|UseHesse|              & Use HESSE error matrix calculation\\
      \verb|NumberOfMinimizations| & Number of minimisation steps\\
      \verb|ErrDef|                & The error definition used in MINOS\\
      \verb|NumberPulls|           & Number of pull fits\\
      \verb|GetContours|           & Determine 2D uncertainty contours \\  % New!
      \verb|UseSimAnnBefore|       & Use Simulated Annealing algorithm directly after \\
                                   & tree-level estimation \\ % New!
      \verb|UseSimAnnWhile|        & Use Simulated Annealing algorithm between global \\
                                   & fits for  \verb|NumberOfMinimizations| $>1$ \\ % New!
      \verb|TempRedSimAnn|         & Temperature reduction factor in Simulated Annealing \\ % New!
      \verb|MaxCallsSimAnn|        & Maximum number of calls in Simulated Annealing \\ % New!
      \verb|InitTempSimAnn|        & Initial temperature of the Simulated Annealing \\ % New!
      \verb|Verbose|               & Verbose mode \\
      \verb|ScanParameters|        & Scan one- or two-dimensional parameter space and write \\
                                   & $\chi^2$-surface to ROOT file \\
      \verb|scanParameter|         & Specify parameters to scan, the scan range and the  \\
                                   & number of scan steps for \verb|ScanParameters| \\
      \verb|PerformFit|            & Controls whether fit is performed or not \\
      \verb|RandomGeneratorSeed|   & Seed for the random number generator \\
      \verb|MaxCalculatorTime|     & Maximal allowed time for the SUSY calculator code \\
      \hline
    \end{tabular}
  \end{center}
\end{table}

\paragraph*{Keys} 
The commands in this group act on the contents of the line after the
key. Available keys are:
\begin{itemize}
\item \verb|#|\\ 
  Comment line. Everything after \verb|#| is ignored by Fittino.
\item \verb|nofit <observable> <value> +- <uncertainty>|\\
  This key can be used to specify observables which shall only be used
  in the step of the initialisation of parameters using tree-level
  relations. Typical examples are the chargino mixing angles
  $\cos2\phi_L$ and $\cos2\phi_R$.
\end{itemize}

\paragraph*{Observables} 
The commands in this group specify the measured (or simulated)
observables and their values and uncertainties. The following types of
observables are available:
\begin{itemize}
\item \verb|mass<name> <value> +- <uncertainty> [ +- <theo_uncertainty> ]|\\
  Specifies the mass of the particle \verb|<name>|. All available
  particles are listed in Tab.~\ref{tab:fittino_particles}. If no
  uncertainty is given, the particle mass is not used in the $\chi^2$
  of the fit. If more than one uncertainty is given, the uncertainties
  are added in quadrature. This is useful in case large theoretical
  uncertainties on the predicted particle mass exist. For the time
  being it is assumed that treating the theoretical uncertainties as
  uncorrelated with the experimental error and a Gaussian
  probability distribution is reasonable.
  
  If no unit is given in \verb|<value>| and \verb|<uncertainty>|, they
  are assumed to be given in GeV. Other supported units range from eV
  to EeV.
  \begin{table}[tp]
  \caption[Particles known to Fittino]{\sl Particles known to Fittino.
  The following particles can be used after \texttt{mass}, \texttt{sigma}, 
  \texttt{BR} and \texttt{width}. Antiparticles are identified by 
  \texttt{\~} after the particle name.}\label{tab:fittino_particles}
  \begin{center}
    \begin{tabular}{|l|l||l|l|}
      \hline
      Particle Name & Explanation & Particle Name & Explanation \\
      \hline\hline
      \verb|W|           & W boson                               &     \verb|SelectronL|  & $\tilde{\mathrm{e}}_L$ selectron \\  
      \verb|Z|           & Z boson                              &     \verb|SelectronR|  & $\tilde{\mathrm{e}}_R$ selectron \\
      \verb|gamma|       & $\gamma$                               &     \verb|SnueL|       & $\tilde{\nu}_{\mathrm{e}_L}$ electron sneutrino\\
      \verb|gluon|       & gluon                                  &     \verb|SmuL|        & $\tilde{\mu}_L$ smuon\\                    
      \verb|Up|          & u quark                                &     \verb|SmuR|        & $\tilde{\mu}_R$ smuon\\                    
      \verb|Down|        & d quark                                &     \verb|SnumuL|      & $\tilde{\nu}_{\mu_L}$ muon sneutrino\\     
      \verb|Charm|       & c quark                                 &     \verb|Stau1|       & $\tilde{\tau}_L$ stau\\                   
      \verb|Strange|     & s quark                                 &     \verb|Stau2|       & $\tilde{\tau}_R$ stau\\                   
      \verb|Top|         & t quark                                 &     \verb|Snutau1|     & $\tilde{\nu}_{\tau_1}$ tau sneutrino\\    
      \verb|Bottom|      & b quark                                  &     \verb|Gluino|      & $\tilde{\mathrm{g}}$ gluino\\            
      \verb|Electron|    & electron                                 &     \verb|Neutralino1| & $\chi^0_1$ neutralino \\                 
      \verb|Nue|         & electron neutrino $\nu_{\mathrm{e}}$     &     \verb|Neutralino2| & $\chi^0_2$ neutralino \\                 
      \verb|Mu|          & muon                                     &     \verb|Neutralino3| & $\chi^0_3$ neutralino  \\                
      \verb|Numu|        & muon neutrino $\nu_{\mu}$                &     \verb|Neutralino4| & $\chi^0_4$ neutralino  \\                
      \verb|Tau|         & tau                                      &     \verb|Chargino1|   & $\chi^{+}_1$ chargino\\                  
      \verb|Nutau|       & tau neutrino $\nu_{\tau}$                &     \verb|Chargino2|   & $\chi^{+}_2$ chargino\\                  
      \verb|h0|          & h CP-even Higgs boson                 & & \\
      \verb|A0|          & A CP-odd Higgs boson & &  \\
      \verb|H0|          & H CP-even Higgs boson & &  \\
      \verb|Hplus|       & H$^{\pm}$ charged Higgs boson & &  \\
      \verb|SdownL|      & $\tilde{\mathrm{d}}_L$ squark & &  \\
      \verb|SdownR|      & $\tilde{\mathrm{d}}_R$ squark & &  \\
      \verb|SupL|        & $\tilde{\mathrm{u}}_L$ squark & &  \\
      \verb|SupR|        & $\tilde{\mathrm{u}}_R$ squark & &  \\
      \verb|SstrangeL|   & $\tilde{\mathrm{s}}_L$ squark & &  \\
      \verb|SstrangeR|   & $\tilde{\mathrm{s}}_R$ squark & &  \\
      \verb|ScharmL|     & $\tilde{\mathrm{c}}_L$ squark & &  \\
      \verb|ScharmR|     & $\tilde{\mathrm{c}}_R$ squark & &  \\
      \verb|Sbottom1|    & $\tilde{\mathrm{b}}_1$ squark & &  \\
      \verb|Sbottom2|    & $\tilde{\mathrm{b}}_2$ squark & &  \\
      \verb|Stop1|       & $\tilde{\mathrm{t}}_1$ squark & &  \\
      \verb|Stop2|       & $\tilde{\mathrm{t}}_2$ squark & &  \\
      \hline
    \end{tabular}
  \end{center}
\end{table}
\item \verb|edge <type> <mass1> <mass2> [more masses] <value> +-|\\\verb|<uncertainties> alias <alias>|\\
%  \texttt{edge \textless{}alias\textgreater{} \textless{}type\textgreater{} \textless{}observable1\textgreater{} \textless{}observable2\textgreater{} [more observables] \textless{}value\textgreater{} +- \textless{}uncertainty\textgreater{}
%    [ +- \textless{}theo_uncertainty\textgreater{} ]}\\
  Specifies the edge in a mass spectrum. Since SUSY particles tend to
  decay in cascade decays, the masses of intermediate particles can
  often be reconstructed from edges in mass spectra. If the edges are
  transformed into masses before the fit is made, the correlations
  among the reconstructed masses from one spectrum have to be
  specified using \verb|correlationCoefficient|. The command
  \verb|edge| offers the simpler and more straightforward possibility
  to use the edge positions in the mass spectra directly in the fit.
  Momentarily, the following types \verb|<type>| are available:\\
  \begin{tabular}{lll}
    1 & $m_1+m_2$ & \\
    2 & $|m_1-m_2|$ & \\
    3 & $\frac{(m_2^2-m_3^2)(m_3^2-m_1^2)}{m_3^2}$  & \cite{lhcilc}\\   
    4 & $\frac{(m_4^2-m_2^2)(m_3^2-m_1^2)}{m_3^2}$  & \cite{lhcilc}\\ 
    5 & $[(m_4^2+m_2^2)(m_2^2-m_3^2)(m_3^2-m_1^2)$ & \\
      & $-(m_4^2-m_2^2)\sqrt{(m_2^2+m_3^2)^2(m_3^2+m_1^2)^2-16m_2^2m_3^4m_1^2}$ & \\
      & $+ 2m_3^2(m_4^2-m_2^2)(m_2^2-m_1^2)]/(4m_3^2m_2^2)$  & \cite{lhcilc}\\ 
  \end{tabular}\\
  The list of formulae can be easily extended. Here and in most of the
  following commands \verb|<alias>| is an integer number which can be
  used to unambigously identify the input if it is used in other
  commands, such as \verb|correlationCoefficient|.
\item \verb|sigma ( <initial_state> -> <final_state_particles>,<Ecms>,<polarisation1>,|\\
  \verb|<polarisation2>) [ <value> +- <uncertainties> ] alias <alias>|\\
  Specifies the cross section of a given process
  \verb|<initial_state>| \verb|->| \verb|<final_state_particles>|,
  where \verb|<final_state_particles>| is a list of particle names.
  Antiparticles are specified by \verb|<particle>~|. Momentarily only
  $\ee$ processes are implemented, identified by
  \verb|<initial_state>|=\texttt{ee}. The centre-of-mass energy (if no
  unit is given, GeV is assumed) is given by \verb|<Ecms>|, the
  polarisation of the incoming particles are given by
  \verb|<polarisation1>| and \verb|<polarisation2>|.  If no unit of
  the cross-section and the uncertainty is given, they are assumed to
  be given in fb. The integer number \verb|<alias>| identifies the
  cross-section in the \verb|correlationCoefficient| commands. No
  value and uncertainty must be given if the cross section is not used
  as an observable on its own but within \verb|xsbr| instead. 
\item \verb|BR ( <decaying_particle> -> <decay_products> ) [ <value>|\\ \verb|+- <uncertainties> ] alias <alias>|\\
  Specifies the branching fraction of a given particle
  \verb|<decaying_particle>| into the decay products
  \verb|<decay_products>|. The integer number \verb|<alias>|
  identifies the branching fraction in the
  \verb|correlationCoefficient| commands. No value and uncertainty
  must be given if the branching fraction is not used as an observable on
  its own but within \verb|brsum|, \verb|xsbr| or \verb|brratio| instead.
\item \verb|brsum ( br_<alias1> br_<alias2> [...] ) <value>|\\ \verb|+- <uncertainties> alias <alias>|\\
  Specifies the sum of several branching fractions, which all have to
  be defined with a unique alias number before. No value and uncertainty
  must be given if the sum of branching fractions is not used as an observable on
  its own but within \verb|xsbr| or \verb|brratio| instead.
\item \verb|width <particle> <value> +- <uncertainties> alias <alias>|\\
  Specifies the total width of the particle \verb|<particle>|. If no unit is
  given in \verb|<value>| and \verb|<uncertainty>|, they are assumed
  to be given in GeV.
\item \verb|limit mass<name> <|$|$\verb|> <limit>|\\
  Allows the user to specify upper or lower mass limits of yet
  undiscovered SUSY particles. The limit is used in the calculation of
  the $\chi^2$ of the fit in the following way: If the predicted mass
  is in agreement with the limit, the contribution of this observable
  to the total $\chi^2$ is zero. If the limit is violated by the
  predicted mass $m_p$, the $\chi^2$ contribution is
  ((\verb|<limit>|$-m_p$)/(\verb|<limit>|/10))$^2$, i.~e. a 10\,\%
  violation of the limit adds 1 to the $\chi^2$ of the fit.
\item \verb|LEObs ( <name> ) <value> +- <uncertainties> alias <alias>|\\
  Low-energy precision observable. In Fittino~1.1 the following observables 
  names are implemented: 
  \begin{itemize}
  \item \texttt{bsg}: BR($b\ra s\gamma$)
  \item \texttt{gmin2}: $(g-2)_{\mu}$
  \item \texttt{drho}: $\Delta\rho$
  \end{itemize} 
\item \verb|xsbr ( sigma_<alias1> [sigma_<alias2> [...]] br_<alias3>|\\ \verb|[br_<alias4> [...]] )  <value> +- <uncertainties> alias <alias>|\\
  Product of an arbitrary number of cross sections and branching
  fractions. This command must be used in the input file after all
  cross sections and branching fractions have been defined. The product
  of cross sections and branching fractions is then specified by the
  alias numbers of the individual observables. These are automatically
  \emph{not} used in the fit anymore. Therefore the actual values
  entered for the individual observables are irrelevant, if they are
  used in \verb|xsbr|. Instead of \verb|br_<alias>| also 
  \verb|brsum_<alias>| can be used. 
\item \verb|brratio ( br_<alias1> br_<alias2> ) <value> +- <uncertainties> alias <alias>|\\
  Ratio of two branching fractions:
  \verb|br_<alias1>|/\verb|br_<alias2>|. Also here the individual
  branching fractions must be defined before their use in
  \verb|brratio| and are \emph{not} used in the fit. Instead of \verb|br_<alias>| also 
  \verb|brsum_<alias>| can be used.
\item Other observables dedicated for special cases are
  \begin{itemize}
  \item \verb|sin2thetaW <value> +- <uncertainties>|\\
    Specifies the value of $\sin^2\theta_W$ 
  \item \verb|cos2phiR <value> +- <uncertainties>|\\
  \item \verb|cos2phiL <value> +- <uncertainties>|\\
    Specify the values of the chargino mixing angles, used for
    initialisation.
  \end{itemize}
\end{itemize}

\paragraph*{Correlations among observables} After all observables have 
been specified, the following command can be used to specify the
correlation among observables. 
\begin{itemize}
\item \verb|correlationCoefficient <observable1> <observable2> <value>|\\
  If the observables are masses, they are identified by
  \verb|mass<name>|. All other observables are identified by their
  alias number, such as \verb|sigma_<alias>|.
\end{itemize}

\paragraph*{Parameters} Fittino can fit any 
combination of the MSSM-24 and SM parameters given in
Tab.~\ref{tab:fittino_parameters} to the observables given in
the input file. As alternatives to MSSM-24, high-scale
parameters of the mSUGRA, GMSB and AMSB models can be determined. 
The following commands can be used to specify the
parameters that are fitted to the observables and the parameters which
are kept fixed.

\begin{table}[tp]
  \caption[MSSM parameters known to Fittino]{\sl MSSM parameters known to Fittino.
    The following parameters can be used with \texttt{fitParameter}, \texttt{fixParameter} and 
    \texttt{universality}.
  }\label{tab:fittino_parameters}
  \begin{center}
    \begin{tabular}{|l|l|}
      \hline
      Parameter Name & Explanation \\
      \hline\hline
      \verb|TanBeta|      &    Ratio of Higgs vacuum expectation values  \\ 
      \verb|Mu|           &     $\mu$ parameter, controls Higgsino mixing \\
      \verb|Xtau|         &     Tau mixing parameter \\
      \verb|Xtop|         &    Top mixing parameter \\
      \verb|Xbottom|      &    Bottom mixing parameter \\
      \verb|MSelectronR|  &      Right scalar electron mass parameter      \\
      \verb|MSmuR|        &      Right scalar muon mass parameter          \\
      \verb|MStauR|       &       Right scalar tau mass parameter          \\
      \verb|MSelectronL|  &    Left 1st.\ gen.\ scalar lepton mass parameter        \\
      \verb|MSmuL|        &    Left 2nd.\ gen.\ scalar lepton mass parameter   \\
      \verb|MStauL|       &    Left 3rd.\ gen.\ scalar lepton mass parameter   \\
      \verb|MSdownR|      &     Right scalar down mass parameter        \\
      \verb|MSstrangeR|   &     Right scalar strange mass parameter     \\
      \verb|MSbottomR|    &     Right scalar bottom mass parameter      \\
      \verb|MSupR|        &     Right scalar up mass parameter                 \\
      \verb|MScharmR|     &     Right scalar charm mass parameter              \\
      \verb|MStopR|       &     Right scalar top mass parameter                \\
      \verb|MSupL|        &    Left 1st.\ gen.\ scalar quark mass parameter    \\
      \verb|MScharmL|     &    Left 2nd.\ gen.\ scalar quark mass parameter    \\
      \verb|MStopL|       &    Left 3rd.\ gen.\ scalar quark mass parameter    \\
      \verb|M1|           &    $U(1)_Y$ gaugino (Bino) mass parameter         \\
      \verb|M2|           &    $SU(2)_L$ gaugino (Wino) mass parameter        \\
      \verb|M3|           &    $SU(3)_C$ gaugino (gluino) mass parameter      \\
      \verb|massA0|       &    Pseudoscalar Higgs mass     \\
      \verb|massW|        &    W boson mass \\
      \verb|massZ|        &    Z boson mass \\
      \verb|massTop|      &    Top quark mass $m_{\mathrm{t}}(m_{\mathrm{t}})$\\
      \verb|massBottom|   &    Bottom quark mass $m_{\mathrm{b}}(m_{\mathrm{b}})$\\
      \verb|massCharm|    &    Charm quark mass $m_{\mathrm{c}}(m_{\mathrm{c}})$ \\    
      \hline
    \end{tabular}
  \end{center}
\end{table}

\begin{itemize}
\item \verb|fitParameter <parameter> [ <value> [ +- <uncertainty>] ]|\\
  Specifies one of the parameters that should be fitted. The names
  which are available for \verb|<parameter>| are listed in
  Tab.~\ref{tab:fittino_parameters}. If a value \verb|<value>| is
  given, then using \verb|FitAllDirectly| it is possible to specify
  that \verb|<value>| should be the initial value of the parameter in
  the fit. If additionally an uncertainty is given, then this can be
  used in the pull distribution calculation with \verb|CalcPullDist|
  or \verb|CalcIndChisqContr|.
  If no unit for \verb|<value>| is given for a dimension-full
  parameter, then it is assumed to be given in GeV.
\item \verb|fixParameter <parameter> <value>|\\
  Specifies a parameter kept fixed during the fit.
\item \verb|universality <parameter1> <parameter2> [...] |\\
  Specifies that \verb|<parameter2>| (and possible other following
  parameters) should not be fitted on its own, but that it should be
  set to the value of \verb|<parameter1>| during the fit. This is
  useful if unification among generations shall be assumed.
\end{itemize}

\paragraph*{Flags} The following flags can be used to control the 
behaviour of Fittino during the fit and to specify what operations
Fittino should perform.
\begin{itemize}
\item \texttt{fitModel <model>}\\
  SUSY model under study. The default value is \texttt{MSSM}, with the
  24 low-scale SUSY Lagrangian parameters from
  Tab.~\ref{tab:fittino_parameters} as parameters. If
  \texttt{fitModel} is \texttt{mSUGRA}, the mSUGRA parameters
  \texttt{TanBeta}, \texttt{M0}, \texttt{M12} and \texttt{A0} can be
  fitted using \texttt{fitParameter}, and \texttt{SignMu} can be fixed
  with \texttt{fixParameter}. Thus the high-scale parameters can be
  fitted directly to the observables. Correspondingly GMSB and AMSB
  parameters can be fitted/fixed, if  \texttt{fitModel} is set to \texttt{GMSB}
  and to \texttt{AMSB} respectively. In these cases the parameter
  names are  \texttt{TanBeta}, \texttt{Lambda}, \texttt{Mmess}, \texttt{cGrav},
  \texttt{N5} and \texttt{SignMu} in case of \texttt{GMSB} and
  \texttt{TanBeta}, \texttt{M0}, \texttt{M32} and \texttt{SignMu} for
  AMSB models.
\item \texttt{LoopCorrections on|off}\\
  If \texttt{LoopCorrections} is \texttt{off}, then no fit is
  performed but just the tree-level estimates of the parameters are
  calculated. By default it is \texttt{on}.
\item \texttt{ISR on|off}\\
  Switches ISR corrections in the cross-section calculations
  \texttt{on} or \texttt{off}. By default it is \texttt{on}.
\item \texttt{UseGivenStartValues on|off}\\
  If \texttt{UseGivenStartValues} is \texttt{on}, then the start
  values of the parameters in the fit are not determined from
  tree-level estimates, but from the values given in
  \texttt{fitParameter}. By default it is \texttt{off}.
\item \texttt{FitAllDirectly on|off}\\
  If \texttt{FitAllDirectly} is \texttt{on}, then the initial fits of
  subsets of the parameter space (as described above in
  Section~\ref{sec:fittino_itself}) are omitted. By default it is
  \texttt{off}.
\item \texttt{CalcPullDist on|off}\\
  If \texttt{CalcPullDist} is \texttt{on}, then pull distributions for
  all parameters specified with \texttt{fitParameter} are calculated.
  It is necessary that each parameter is given with its value and
  uncertainty. The pull distribution is then calculated with respect
  to the parameter value and the width is compared with the parameter
  uncertainty. For each parameter, a ROOT~\cite{Brun:1997pa} histogram
  is created in the output file \texttt{PullDistributions.root}.
  Additionally, a ROOT tree with all fitted parameters and the
  corresponding $\chi^2$ values is created, which allows for the
  extraction of the parameter correlations. The number of fits per
  Fittino run can be specified using the command \texttt{NumberPulls}.
  By default \texttt{CalcPullDist} is \texttt{off}. This command is
  very useful to test the fitted parameters and their uncertainties
  calculated in a previous run of Fittino.
\item \texttt{CalcIndChisqContr on|off}\\
  If \texttt{CalcIndChisqContr} is \texttt{on}, for each parameter
  specified with \texttt{fitParameter} the individual contribution of
  each observable to the $\Delta\chi^2$ of the fit is calculated, if
  the parameter is varied by $\pm1\,\sigma$.  It is necessary that
  each parameter is given with its value and uncertainty. The
  parameter is varied once by $+1\,\sigma$ and once by $-1\,\sigma$.
  The resulting total $\Delta\chi^2$ and the individual
  $\Delta\chi^2_i$ of each observable $O_i$ are averaged. The total
  $\Delta\chi^2$ indicates the correlation of the parameter with all other
  parameters. If the parameters are not too strongly correlated, the
  individual $\Delta\chi^2_i$ provide a measure for the
  contribution of observable $O_i$ to the determination of the
  parameter. The output is given in the file
  \texttt{fittino\_individual\_chisq\_contr.out}. By default,
  \texttt{CalcIndChisqContr} is \texttt{off}.
\item \texttt{BoundsOnX on|off} \\
  If \texttt{BoundsOnX} is \texttt{on}, then the parameters
  $X_{\tau}$, $X_{\mathrm{t}}$ and $X_{\mathrm{b}}$ are bounded
  between \verb|Xscanlow|$<X<$\verb|Xscanhigh|. By default
  \texttt{BoundsOnX} is \texttt{on}.
\item \texttt{SepFitmA on|off} \\
  If \verb|SepFitmA| is \texttt{on}, a separate fit of 
  $m_{\mathrm{A}_{\mathrm{run}}}$ to $m_{\mathrm{A}_{\mathrm{pole}}}$
  is performed, in order to improve the start value of the parameter
  $m_{\mathrm{A}_{\mathrm{run}}}$. This is necessary in some
  scenarios, where the difference between
  $m_{\mathrm{A}_{\mathrm{pole}}}$ (which is used as a tree-level
  estimate of $m_{\mathrm{A}_{\mathrm{run}}}$) and
  $m_{\mathrm{A}_{\mathrm{run}}}$ is large and correlations of
  $m_{\mathrm{A}_{\mathrm{run}}}$ are strong. By default
  \verb|SepFitmA| is \texttt{off}.
\item \texttt{SepFitTanbMu on|off} \\
  If \verb|SepFitTanbMu| is \texttt{on}, a separate fit of $\tanb$
  and $\mu$ is performed before the main fit and directly after the
  separate fit of $m_{\mathrm{A}_{\mathrm{run}}}$ to
  $m_{\mathrm{A}_{\mathrm{pole}}}$. Only gaugino observables are used.
  By default \verb|SepFitTanbMu| is \texttt{off}.
\item \texttt{ScanX on|off}  \\
  If \verb|ScanX| is \texttt{on}, then the parameters
  $X_{\mathrm{t}}$ and $X_{\mathrm{b}}$ are individually
  scanned in the range $-6000<X<2000$ before the main fit and before
  the separate fit of the squark sector. This helps to avoid local
  minima which typically occur at parameter values with the wrong
  sign. By default \verb|ScanX| is \texttt{on}.
\item \texttt{SepFitTanbX on|off} \\
  If \verb|SepFitTanbX| is \texttt{on}, a separate fit of $\tanb$,
  $M_{\tilde{\mathrm{t}}_R}$, $M_{\tilde{\mathrm{t}}_L}$,
  $X_{\mathrm{t}}$ and $X_{\mathrm{b}}$ is performed before the
  main fit and after the separate fit of the squark sector. Only
  squark sector observables are used. By default \verb|SepFitTanbX| is
  \texttt{on}.
\item \verb|Calculator <calculator_name> <path>|\\
  Specifies the tool for the calculation of the theory predictions.
  \verb|<path>| specifies the location of the calculator executable.
  Currently SPheno is implemented, but also any other tool capable of
  input and output according to the SUSY Les Houches
  Accord~\cite{Skands:2003cj} can be easily interfaced with Fittino.
\item \texttt{UseMinos on|off}\\
  If \texttt{UseMinos} is \texttt{on}, then MINOS is used to perform a
  detailed uncertainty analysis after MINIMIZE converged.
  \texttt{UseMinos} implies automatically that \texttt{UseHesse} is
  \texttt{on} (see below).  Since this can take very long (order of
  several days on a PIII 1.3GHz in a typical fit of the full MSSM
  spectrum), this option is \texttt{off} by default.
\item \texttt{UseHesse on|off}\\
  If \texttt{UseHesse} is \texttt{on}, then the HESSE function in
  MINUIT is used to perform a detailed error matrix calculation after
  MINIMIZE converged, assuming parabolic errors. By default,
  \texttt{UseHesse} is \texttt{off}.
\item \texttt{GetContours on|off}\\
  If \texttt{GetContours} is \texttt{on}, after MINOS the
  two-dimensional uncertainty contours of all combinations of two
  parameters are calculated and stored in the output files
  \texttt{FitContours.root} (fit contours with an error definition of
  $\Delta\chi^2=1$ (thus the resulting two-dimensional surface will
  not be the two-dimensional 68\,\% region) and all parameters free)
  and \texttt{FitContours\_2params\_free.root} (just two parameters
  free per contour). Since this can take very long (order of several
  days to several weeks in a typical fit of the full MSSM spectrum),
  this option is \texttt{off} by default.
\item \texttt{UseSimAnnBefore on|off}\\
  If \texttt{UseSimAnnBefore} is \texttt{on}, a Simulated Annealing
  algorithm is used directly after the tree-level estimates of the
  parameters. Typically this algorithm is very robust and finds the
  global minimum of the $\chi^2$ reliably. The maximum number of calls
  to the theory code in the Simulated Annealing process is regulated
  using \texttt{MaxCallsSimAnn}, which by default is 300\,000. The
  temperature reduction can be set using \texttt{TempRedSimAnn}, which
  by default is 0.4.  The history of the Simulated Annealing process
  is recorded in a ROOT ntuple the file \texttt{SimAnnNtupFile.root}.
  By default, \texttt{UseSimAnnBefore} is \texttt{off}.
\item \texttt{UseSimAnnWhile on|off}\\
  If \texttt{UseSimAnnWhile} is \texttt{on} and
  \texttt{NumberOfMinimizations} is larger than 1, Simulated Annealing
  is used in between subsequent calls to MINIMIZE in the global fit.
  It works as explained for \texttt{UseSimAnnBefore}. 
  By default, \texttt{UseSimAnnWhile} is \texttt{off}.
\item \verb|NumberOfMinimizations <number>|\\
  In very complex cases the first call to MINIMIZE often does converge
  near the true minimum of the fit, but the convergence criteria are
  often not fulfilled after the first call to MINIMIZE. Therefore,
  MINIMIZE can be called \verb|<number>| times after each other. By
  default, \texttt{NumberOfMinimizations} is 1.
\item  \verb|ErrDef <real_number>|\\
  If the parameter space of the fit is very complex, either due to a
  large number of d.o.f. or because of observables with large
  uncertainties, sometimes MINOS is not able to find a positive
  definite error matrix with the standard setting of $\Delta\chi^2=1$
  for the definition of the $1\,\sigma$ bound of the parameters.
  Therefore, using \texttt{ErrDef}, the error definition can be
  changed from 1 to any other positive number. After MINOS is finished,
  the uncertainties and the covariance matrix found by MINOS is
  re-transformed assuming parabolic errors such as to represent an
  error definition of 1. With small error definitions, MINOS finds the
  uncertainties more easily, since it is more seldom trapped in local
  minima close to the absolute minimum. On the other hand, the
  uncertainties on the parameters are less precise for small error
  definitions. By default, \texttt{ErrDef} is set to 1.
\item \verb|NumberPulls <number>|\\
  Specifies the number of individual fits for the calculation of pull
  distributions in one run of Fittino. For each fit, the observables
  are smeared in a Gaussian form according to their covariance matrix.
  The initialisation of the random number generator used for the
  smearing uses either the seed provided with
  \verb|RandomGeneratorSeed| or the system time in seconds plus the
  system uptime in seconds plus the PID of the Fittino process plus
  the amount of currently available free swap space. Thus it is highly
  improbable that two Fittino processes share the same initialization.
  By default, \texttt{NumberPulls} is set to 10. Only MINIMIZE is used
  in the fit, MINOS is switched off.
\item \verb|TempRedSimAnn <number>|\\
  Temperature reduction factor for the Simulated Annealing algorithm.
  Should be in the interval $]0,1[$. The default value is 0.4.
\item \verb|MaxCallsSimAnn <number>|\\
  Maximum number of calls to the theory code in one Simulated
  Annealing run. The default value is 300\,000.
\item \verb|InitTempSimAnn <number>|\\
  Initial temperature of the Similated Annealing algorithm. If it is
  not set or negative, the initial temperature is calculated from the
  variation of the $\chi^2$ at the starting point.
\item \texttt{Verbose on|off}\\
  If \texttt{Verbose} is \texttt{on}, all parameter and observable
  values are printed to \texttt{stdout} for every iteration of the
  fit. If \texttt{Verbose} is \texttt{off}, only every tenth iteration
  is shown. By default, \texttt{Verbose} is \texttt{on}.
\item \texttt{ScanParameters on|off}\\
  If \texttt{ScanParameters} is \texttt{on}, either a one- or
  two-dimensional scan of the $\chi^2$ surface is performed and stored
  in the ROOT file \texttt{ParameterScan.root}.  At most two
  parameters to be scanned must be specified using
  \texttt{scanParameter}. The scan is performed after the fit. By
  default, \texttt{scanParameters} is \texttt{off}.
\item \verb|scanParameter <parameter> <lbound> <ubound> <nsteps>|\\
  Specifies one parameter to be scanned if \texttt{scanParameters} is
  \texttt{on}. The lower bound \verb|<lbound>| and upper bound
  \verb|<ubound>| are assumed to be in GeV if no unit is given. The
  argument \verb|<nsteps>| specifies the number of steps in this
  dimension of the scan. One or two \texttt{scanParameter} statements
  are needed for a scan.
\item \texttt{PerformFit on|off}\\
  If \texttt{PerformFit} is \texttt{off}, no fit is performed.
  Instead, evaluation tasks like \verb|scanParameters| are executed
  directly. By default \texttt{PerformFit} is \texttt{on}.
\item \verb|RandomGeneratorSeed <value>|\\
  If \verb|RandomGeneratorSeed| is given, a seed \verb|<value>| is
  used for the initialization of the random number generator in
  \verb|CalcPullDist| and in the simulated annealing. If
  \verb|RandomGeneratorSeed| is not given, the seed is calculated from
  a combination of system uptime, free swap space, pid of the Fittino
  process and system time, which should be unique for every run of
  Fittino.
\item \verb|MaxCalculatorTime <value>|\\
  Specified the maximal time in seconds allowed for one calculation of
  the SUSY observables by the calculator code. The default value is 20
  seconds.
\end{itemize}
An example for a Fittino input file \texttt{fittino.in} can be found
in Section~\ref{app:fittino_input_file}.

%=======================================================================
\section{Quickstart Guide}\label{sec:quickstart}

After specifying the observables with the appropriate commands from
Tab.~\ref{tab:fittino_commands}, the parameters to be fitted with
\texttt{fitParameter} and the fixed parameters with
\texttt{fixParameter}, a fit with the following set of flags from
Tab.~\ref{tab:fittino_flags} can be tried:
\begin{verbatim}
LoopCorrections       off
ISR                   on

UseGivenStartValues   on
FitAllDirectly        on
BoundsOnX             off
ScanX                 off
SepFitTanbX           off
SepFitTanbMu          off
SepFitmA              off

UseMinos              off
UseHesse              off
GetContours           off
UseSimAnnBefore       off
UseSimAnnWhile        off

ScanParameters        off
CalcPullDist          off
CalcIndChisqContr     off

Calculator             YOUR_CALCULATOR  /path/to/your/theory/code

NumberOfMinimizations  1
ErrDef                 1.0
\end{verbatim}
With this parameter setting Fittino perfoms no fit but just
calculates the tree-level estimates of the parameters, if in the
Fittino code the theory program \verb|YOUR_CALCULATOR| is known.  If
the code stops because a tree-level estimate could not be made (e.g.{}
``\texttt{Value massSupL not found}''), then additionally the
requested observable should be specified using the \texttt{nofit}
command. This is done in order to give the user the responsibility
over the inputs used for the tree-level estimates. The \texttt{nofit}
command makes sure that the observable is not used in the fit.

After the tree-level estimates have successfully been determined, 
\begin{verbatim}
LoopCorrections       on
\end{verbatim}
should be set.  This ensures that the fit is actually started. No
Simulated Annealing or subsector fit is performed. If the `true'
parameter values are known, for the first try those should be entered
in the \texttt{fitParameter} and \texttt{fixParameter} commands, in
order to ensure that the global fit starts with the true parameter
settings. The $\chi^2$ value printed after the first call to the
theory code should then be close to $0$, if all parameters and
observables are entered correctly. Even if the observables are smeared
within their experimental and theoretical uncertainties, the start of
the fit with the original set of parameters can be helpful to learn
about the behaviour of the system. The reason for this step is that
the start with the `true' parameter settings allows a fast check
whether the system is constrained enough around the $\chi^2$ minimum
to yield a realistic uncertaity estimate. If the resulting uncertainty
matrix in \texttt{fittino.out} is not called \texttt{Error matrix
  accurate}, then the significance of the observables for the
determination of the uncertainties of the parameters at the $\chi^2$
minimum is not strong enough. New observables have to be added, the
observable uncertainties have to be reduced (if feasible) or the
number of parameters has to be decreased. If nothing of the above is
possible, a fit with
\begin{verbatim}
UseMinos              on
\end{verbatim}
may still yield a correct error matrix in many cases. 

After it is ensured that the problem is constrained strongly enough to
yield a meaningful uncertainty matrix, the full fit can be tested.
This can be done either with the subsector fit method or with
Simulated Annealing. In the former case, 
\begin{verbatim}
UseGivenStartValues   off
FitAllDirectly        off
\end{verbatim}
has to be set. If the subsector fits do not converge closely enough to
the true $\chi^2$ minimum of the problem, switching on some of the
additional subsector fits may help, depending on the individual
problem. Large values of the $\chi^2$ in the early phases of the
subsector fits and the subsequent global fit are generally no reason
to worry.

If a more stable (but also slower) way should be tried, the Simulated
Annealing can be used. in this case 
\begin{verbatim}
UseGivenStartValues   off
FitAllDirectly        on
UseSimAnnBefore       on
\end{verbatim}
has to be used. The default setting of 
\begin{verbatim}
TempRedSimAnn         0.4
MaxCallsSimAnn        300000
\end{verbatim}
ensures a stable convergence for most problems . If it is felt that
the convergence could be faster, a smaller value of
\texttt{InitTempSimAnn} than the one chosen by Fittino and maybe a
smaller \texttt{TempRedSimAnn} can be tried. Should no convergence
occur, all values can be increased.

If the fit converges to a minimum close to the expected one instead of
the true minimum (or, in case the true parameter values are unknown,
it is felt that the convergence is not sufficient), Fittino can be
made even more robust by setting \texttt{NumberOfMinimizations} to a
value larger than one (typically 2 or 3), and using
\begin{verbatim}
UseSimAnnWhile       on
\end{verbatim}
This ensure that several subsequent global fits with intermediate
Simulated Annealing algorithms are performed.

After the correct minimum is found and the global fit successfully
converged, the uncertainty matrix can be refined and the asymmetric
uncertainties can be calculated using
\begin{verbatim}
UseMinos              on
\end{verbatim}
After the fitted parameter values and their uncertainties have been
obtained in this way, a further analysis of the parameter
determination can be done using \texttt{GetContours},
\texttt{CalcPullDist} or \texttt{CalcIndChisqContr}. In the case of
\texttt{CalcPullDist}, the parameter uncertainties found by Fittino in
the previous run have to be entered together with the true parameter
values in \texttt{fitParameter}. In the case of
\texttt{CalcIndChisqContr}, both the fitted central parameter values
and the parameter uncertainties as found by Fittino have to be entered
in \texttt{fitParameter}.

%=======================================================================
\section{The Fittino Output Files}\label{app:fittino_out}

Depending on the requested operation, Fittino saves its results in the
following files: 
\begin{itemize}
\item \texttt{fittino.out}: Main output file. It contains the
  observables, their covariance matrix, the fixed and fitted MSSM
  parameters and their correlation and covariance matrices.
  Additionally, information on the $\chi^2$ and the accuracy of the
  error matrix estimate is shown. An example of this output file is
  given in Section~\ref{app:output_file}.
\item \texttt{FitContours.root}: In case a MINOS uncertainty analysis
  has been performed (using \texttt{UseMinos}), the two-dimensional
  uncertainty contours of each pair of parameters are calculated if
  the option \texttt{GetContours} is \texttt{on}. The contours are
  stored in the ROOT~\cite{Brun:1997pa} format in this file.
\item \texttt{FitContours\_2params\_free.root}: As
  \texttt{FitContours.root}, but for each two dimensional contour of
  two parameters all other parameters are fixed to their best fit
  values.
\item \texttt{fittino\_individual\_chisq\_contr.out}: In case that an
  analysis of which observables determine which parameters is
  requested using \texttt{CalcIndChisqContr}, the individual
  contributions to the $\Delta\chi^2$ are listed in this file.
\item \texttt{PullDistributions.root}: In case pull distributions are
  calculated using \texttt{CalcPullDistr}, the pull distributions and
  the distribution of the total $\chi^2$ of the fits are stored in
  this file in the ROOT format. One-dimensional projections of the
  distribution of fitted parameter values are available in ROOT
  histograms. The full correlation of parameters can be deduced from
  the ROOT tree containing all fitted parameter values and the
  corresponding $\chi^2$ values.
\item \texttt{SimAnnNtupFile.root}: This file contains one ROOT-ntuple
  for each run of the Simulated Annealing algorithm. The ntuple
  contains the $\chi^2$ value and the parameter settings at each step.
\item \texttt{ParameterScan.root}: In case a parameter scan is
  performed, a TGraph (or TGraph2D) object describing the $\chi^2$
  curve (plane) is stored in this file.
\end{itemize}
Examples for the contents of the output files are given in the
appendix.

\section{Summary}\label{sec:summary}

Fittino is a versatile program with flexible user interface which is
designed to perform a global $\chi^2$ fit of MSSM and SM parameters to
observables from present and future collider experiments. Using
Fittino, information about the feasibility of a unique reconstruction
of MSSM parameters can be obtained, parameter uncertainties and
correlations for simulated sets of observables can be determined and
it can be used to optimise the observables for a desired parameter
uncertainty. Additionally, the fit results of Fittino can be tested by
the automatic generation of pull distributions and $\chi^2$
distributions. Two-dimensional uncertainty contours can be obtained to
study nonlinearities in the parameter correlations and to visualise
the results. The contribution of the individual observables to the
confinement of each parameter can be obtained by studying the
individual contributions of each observable to the $\Delta\chi^2$ for
parameter variations by $\pm1\sigma$. 

Fittino has been tested using global fits of 19 MSSM parameters and
one SM parameter to observables from the LHC and the ILC. The program
has also been used to directly fit high-scale mSUGRA parameters
to anticipated LHC and ILC measurements. No prior
knowledge of the parameters has been assumed at any step. The
parameters are reconstructed entirely from the observables. It has
been shown that the fit strategy of Fittino, using tree-level
estimates of the parameters, subsector fits in parts of the parameter
space to improve the tree-level estimates and then global fits is
stable even for different MSSM scenarios like SPS1a, SPS1a' and SPS7.
The possibility to include Simulated Annealing in order to find the
global $\chi^2$ minimum before the global fit provides a further
stable method to find the optimal parameter values.

In the future, Fittino can be used for systematic studies of the
dependence on measurement uncertainties. The most crucial observables
for the parameter determination can be identified and the detector and
machine, especially the distribution of the luminosities with different
centre-of-mass energies and beam polarisations, can be optimised for
most precise parameter measurements. Also different sets of SUSY
parameters and their phenomenology can be studied. The influence of
theoretical uncertainties can be determined and regions with need for
improvement can be identified. 

\section*{Acknowledgements}\label{sec:ack}
The authors wish to thank Werner Porod (especially for SPheno), Gudrid
Moortgat-Pick and the whole SPA working group for very fruitful
discussions and lots of help.

%---------------------------------------------------------------------
\appendix
\section{Fittino Input and Output Files}
\subsection{Example for the Fittino Steering File}\label{app:fittino_input_file}

In the following an example for the Fittino input file
\texttt{fittino.in} is given.  For demonstration, it shows a fit of
just two observables, $\tanb$ and $\mu$. All other parameters are kept
fixed. An assorted selection of input observables from the ILC is used
for illustration. The order of the inputs is mostly arbitrary, apart
from the correlations of the observables, which always must be stated
after all observables are given.

\begin{verbatim}
#########################################################################
###                         Fittino input file                        ###
#########################################################################
###                 This is an example steering file.                 ###
###                        P. Bechtle, 20040520                       ###
#########################################################################

# SM observables
massW                    80.3078  GeV +- 0.039  GeV
massZ                    91.1187 GeV +- 0.0021 GeV
massTop                  174.3   GeV +- 0.05    GeV
massBottom               4.2     GeV +- 0.5    GeV
sin2ThetaW               0.23113 +- 0.00015
alphas                   0.1172  +- 0.002

# Higgs sector at ILC 500 and ILC 1000
massh0                   108.67 GeV +- 0.05 GeV +- 0.5 GeV # Degrassi et al
massA0                   311.036 GeV +- 1.3 GeV      # not in LHC
# Gauginos at ILC 500 and ILC 1000
massNeutralino1          95.9804 GeV +- 0.05 GeV # ~chi_10
massNeutralino2          180.397 GeV +- 0.08 GeV # ~chi_20
massChargino1            180.033 GeV +- 0.55 GeV # ~chi_1+
massChargino2            381.929 GeV +- 3.0 GeV  # ~chi_2+
# plus squark and slepton masses, if necessary with preceeding 'nofit'
nofit massSupL           561.539 GeV +- 9.8 GeV  #   ~uL
nofit massSupR           543.35  GeV +- 23.6 GeV #   ~uR
nofit massSdownL         561.54  GeV +- 9.8 GeV  #   ~dL
nofit massSdownR         543.348 GeV +- 23.6 GeV #   ~dR
nofit massSstrangeL      561.54  GeV +- 9.8 GeV  #   ~sL
nofit massSstrangeR      543.348 GeV +- 23.6 GeV #   ~sR
nofit massScharmL        561.54  GeV +- 9.8 GeV  #   ~cL
nofit massScharmR        543.348 GeV +- 23.6 GeV #   ~cR
nofit massSbottom1       502.059 GeV +- 5.7  GeV #   ~b_1       
nofit massSbottom2       541.81  GeV +- 6.2  GeV #   ~b_2       
nofit massStop1          365.819 GeV +- 2.0 GeV  #   ~t_1       
nofit massStop2          600.037 GeV +- 20.0 GeV #   ~t_2       
nofit massSelectronL     190.209 GeV +- 0.2 GeV  #   ~e_L-      
nofit massSelectronR     124.883 GeV +- 0.05 GeV #   ~e_R-      
nofit massSnueL          172.947 GeV +- 0.7 GeV  #   ~nu_eL
nofit massSmuL           190.237 GeV +- 0.5 GeV  #   ~mu_L-     
nofit massSmuR           124.837 GeV +- 0.2 GeV  #   ~mu_R-     
nofit massStau1          107.292 GeV +- 0.3 GeV  #   ~tau_1-
nofit massStau2          195.290 GeV +- 1.1 GeV  #   ~tau_2- 
nofit massGluino         603.639 GeV +- 6.4 GeV  #   ~g 
nofit massNeutralino1    97.7662 GeV +- 0.05 GeV #   ~chi_10
nofit massNeutralino2    184.345 GeV +- 0.08 GeV #   ~chi_20
nofit massNeutralino3    -404.134 GeV +- 4.0 GeV #   ~chi_30
nofit massNeutralino4    417.037 GeV +- 2.3 GeV  #   ~chi_40
nofit massChargino1      184.132 GeV +- 0.55 GeV #   ~chi_1+
nofit massChargino2      418.495 GeV +- 3.0 GeV  #   ~chi_2+

# possible edges
# edge 1 massNeutralino1 massNeutralino2 263.50279 GeV +- 1.2 GeV alias 1
# edge 2 massNeutralino1 massNeutralino2 79.09719  GeV +- 1.2 GeV alias 2

# estimated chargino mixing angles from ILC 500
nofit cos2PhiL                 0.6737 +- 0.05       # rough estimate
nofit cos2PhiR                 0.8978 +- 0.05       # rough estimate

# ILC 500 Cross-sections
sigma ( ee->Neutralino1 Neutralino2,500.,0.8,-0.6 ) 20.5026 fb +- 2.0 fb alias 1
sigma ( ee->Neutralino2 Neutralino2,500.,0.8,-0.6 ) 5.62767 fb +- 2.0 fb alias 2
sigma ( ee->Chargino1 Chargino1~,500.,0.8,-0.6 )    13.6598 fb +- 1.0 fb alias 6
sigma ( ee->Z h0,500., 0.8, -0.6 )               29.0566 fb +- 0.21 fb   alias 7
sigma ( ee->Chargino1 Chargino1~,500.,-0.8,0.6 )  462.321 fb +- 5.0 fb alias 8
sigma ( ee->Neutralino1 Neutralino2,500.,-0.8,0.6 ) 180.84  fb +- 2.0 fb alias 9
sigma ( ee->Neutralino2 Neutralino2,500.,-0.8,0.6 ) 200.396 fb +- 2.0 fb alias 10

# ILC 500 Branching Fractions
# BR ( h0 -> Bottom Bottom~ )                0.824057   +- 0.019      alias 1
# BR ( h0 -> Charm  Charm~ )                 0.0405547  +- 0.02       alias 2
# BR ( h0 -> Tau  Tau~ )                     0.134444   +- 0.02       alias 3

# Product of cross sections and branching fractions:
# xsbr ( sigma_7 br_1 ) 32.874 +- 0.5 alias 1
# Ratio of branching fractions
# brratio ( br_1 br_2 ) 20.6259 +- 0.5 alias 1

# Correlations among observables
# correlationCoefficient   massChargino1   massNeutralino1    0.05

# Parameters to be fitted
fitParameter  TanBeta          10.0     
fitParameter  Mu               358.635 GeV 

# Fixed Parameters
fixParameter  Atau            -3884.46 GeV 
fixParameter  MSelectronR      135.762 GeV 
fixParameter  MStauR           133.564 GeV 
fixParameter  MSelectronL       195.21 GeV 
fixParameter  MStauL           194.277 GeV 
fixParameter  Atop            -506.936 GeV 
fixParameter  Abottom         -4444.56 GeV 
fixParameter  MSdownR          528.135 GeV 
fixParameter  MSbottomR        524.719 GeV 
fixParameter  MSupR            530.244 GeV 
fixParameter  MStopR           424.515 GeV 
fixParameter  MSupL            548.704 GeV 
fixParameter  MStopL           499.986 GeV 
fixParameter  M1               101.814 GeV 
fixParameter  M2               191.771 GeV 
fixParameter  M3               588.798 GeV 
fixParameter  massA0           399.763 GeV
fixParameter  massTop          174.3   GeV
fixParameter  massBottom      4.200e+00   GeV
fixParameter  massCharm       1.2e+00   GeV

# Universalities among parameters
fixParameter  MSmuR          135.76 GeV # 1.35760e+02 GeV #    135.76 GeV
fixParameter  MSmuL          195.21 GeV # 1.95199e+02 GeV #    195.21 GeV
fixParameter  MSstrangeR     528.14 GeV   # 5.28134e+02 GeV # 528.14 GeV
fixParameter  MScharmR       530.253 GeV  # 5.30270e+02 GeV # 530.253 GeV
fixParameter  MScharmL       548.705 GeV  # 5.48702e+02 GeV # 548.705 GeV

# possible unifications
# universality MSelectronR MSmuR
# universality MSelectronL MSmuL
# universality MSdownR MSstrangeR
# universality MSupR MScharmR
# universality MSupL MScharmL

# switches
LoopCorrections       on          # Use full loop corrections in SPHENO
ISR                   on          # Switch on ISR

UseGivenStartValues   on          
FitAllDirectly        on

ScanParameters        off
CalcPullDist          off
CalcIndChisqContr     off

# Calculator             SPHENO  /home/bechtle/programs/SPheno2.2.2/SPheno

UseMinos              on

NumberOfMinimizations  1
ErrDef                 1.
NumberPulls            3
\end{verbatim}

%---------------------------------------------------------------------
%\clearpage
\subsection{Example for the Fittino Output File}\label{app:output_file}

The result of the fit from the input file \texttt{fittino.in}
shown in Section~\ref{app:fittino_input_file} is given in the file
\texttt{fittino.out}. It is presented in the following. The lengthy table
of the correlations of the observables is abbreviated. Please note the
message \texttt{Error Matrix Accurate} in the end of the file. If
this message is not given, the fit uncertainties are not reliable,
since MINUIT could not find a positive definite uncertainty matrix.

\begin{verbatim}
#################### Fittino Fit Summary ####################
created by Fittino version 1.0.3
on Wednesday, August 10, 2005 at 07:28:02

Input values:
=============
               massZ        91.1187 +-       0.0021
               massW        80.3078 +-        0.039
           massCharm            1.2 +-          0.2
          massBottom            4.2 +-          0.5
             massTop          174.3 +-         0.05
             massTau        1.77699 +-      0.00029
              alphas         0.1172 +-        0.002
             alphaem        127.934 +-        0.027
          sin2ThetaW        0.23113 +-      0.00015
              massh0         108.67 +-     0.502494
              massA0        311.036 +-          1.3
     massNeutralino1        95.9804 +-         0.05
     massNeutralino2        180.397 +-         0.08
       massChargino1        180.033 +-         0.55
       massChargino2        381.929 +-            3
ee  -> Neutralino1 Neutralino2       20.502600 +- 2.000000
ee  -> Neutralino2 Neutralino2       5.627670 +- 2.000000
ee  -> Chargino1 Chargino1~       13.659800 +- 1.000000
        ee  -> Z h0         29.0566 +-         0.21
ee  -> Chargino1 Chargino1~       462.321000 +- 5.000000
ee  -> Neutralino1 Neutralino2       180.840000 +- 2.000000
ee  -> Neutralino2 Neutralino2       200.396000 +- 2.000000

Covariance matrix for input value:
==================================

                                    massh0      massA0    all observables ...
              massh0              0.252500    0.000000    ...
              massA0              0.000000    1.690000    ...
                 .                    .           .
           all observables            .           .
                 .                    .           .
                 .                    .           .
                 .                    .           .

Fixed values: 
=============
                Xtau       -3884.46
         MSelectronR        135.762
              MStauR        133.564
         MSelectronL         195.21
              MStauL        194.277
                Xtop       -506.936
             Xbottom       -4444.56
             MSdownR        528.135
           MSbottomR        524.719
               MSupR        530.244
              MStopR        424.515
               MSupL        548.704
              MStopL        499.986
                  M1        101.814
                  M2        191.771
                  M3        588.798
              massA0        399.763
               MSmuR         135.76
               MSmuL         195.21
          MSstrangeR         528.14
            MScharmR        530.253
            MScharmL        548.705
             massTop          174.3
          massBottom            4.2
           massCharm            1.2

Fitted values:
==============
             TanBeta        9.99992 +-    0.0935112 
                  Mu        358.639 +-      1.00434 

Covariance matrix for fitted parameters:
========================================

                        TanBeta          Mu
             TanBeta   0.00874435  -0.0532778 
                  Mu   -0.0532778     1.00871 
 
Correlation matrix for fitted parameters:
=========================================

                        TanBeta          Mu
             TanBeta            1   -0.567283 
                  Mu    -0.567283           1 
 
Chisq of the fit:
=================

                   chisq = 0.000007

Status of the minimization:
===========================

                   Error Matrix accurate

##################### End of Fit Summary ####################
\end{verbatim}
%\clearpage
\subsection{Examples for the Two-Dimensional Uncertainty Contour Plots}\label{sec:uncert}

\begin{figure}[t]
%  \placefigs{figures/2Param_Fit_Contour_of_Mu_vs_TanBeta.eps}{0.49}
%  \placefigs{figures/2Param_Fit_Contour_of_M3_vs_TanBeta.eps}{0.49}
%  \placefigs{figures/2Param_Fit_Contour_of_MSdownR_vs_MSelectronR.eps}{0.49}
%  \placefigs{figures/2Param_Fit_Contour_of_MSupR_vs_MSelectronR.eps}{0.49}
  \placefigs{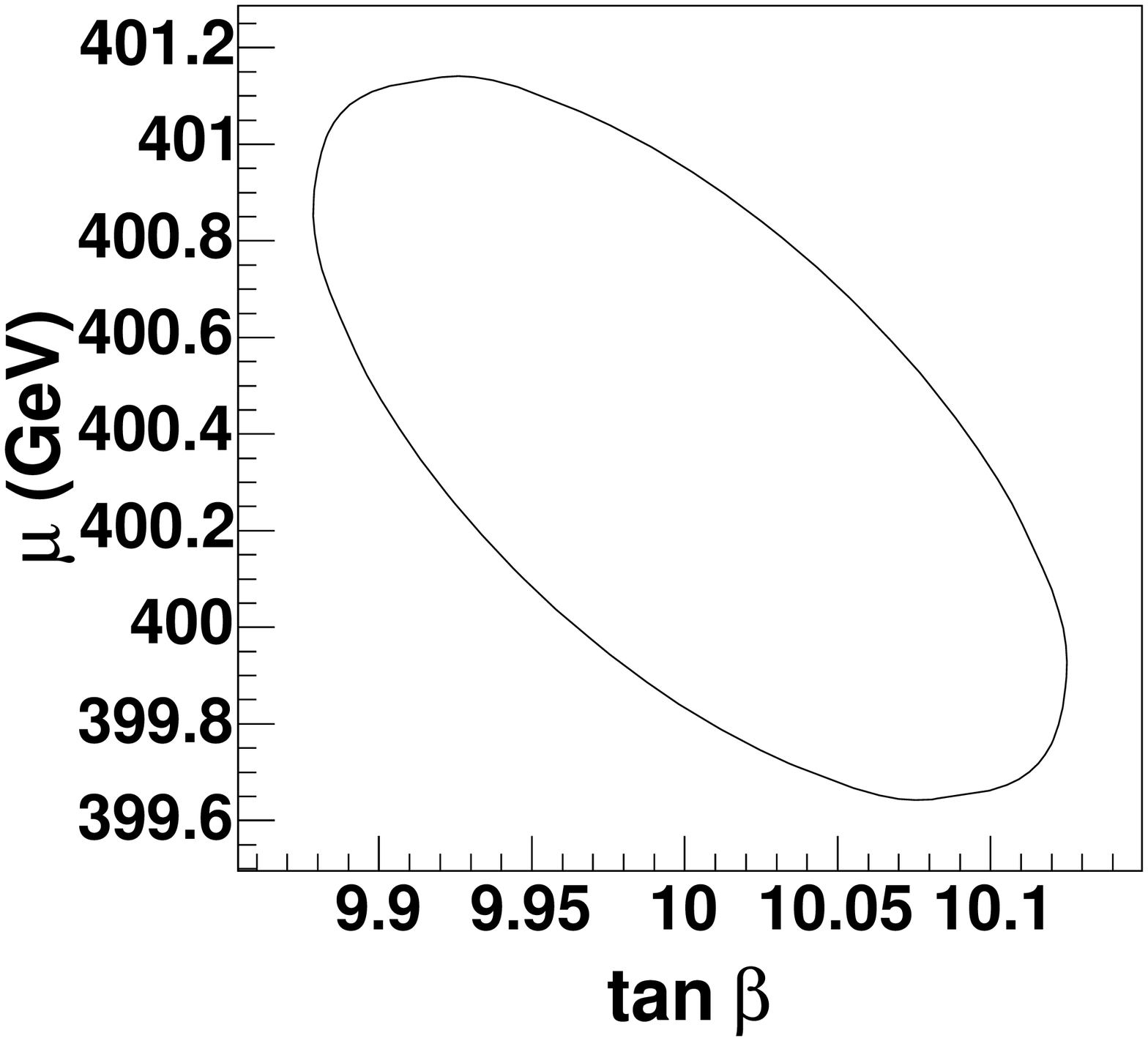}{0.49}
  \placefigs{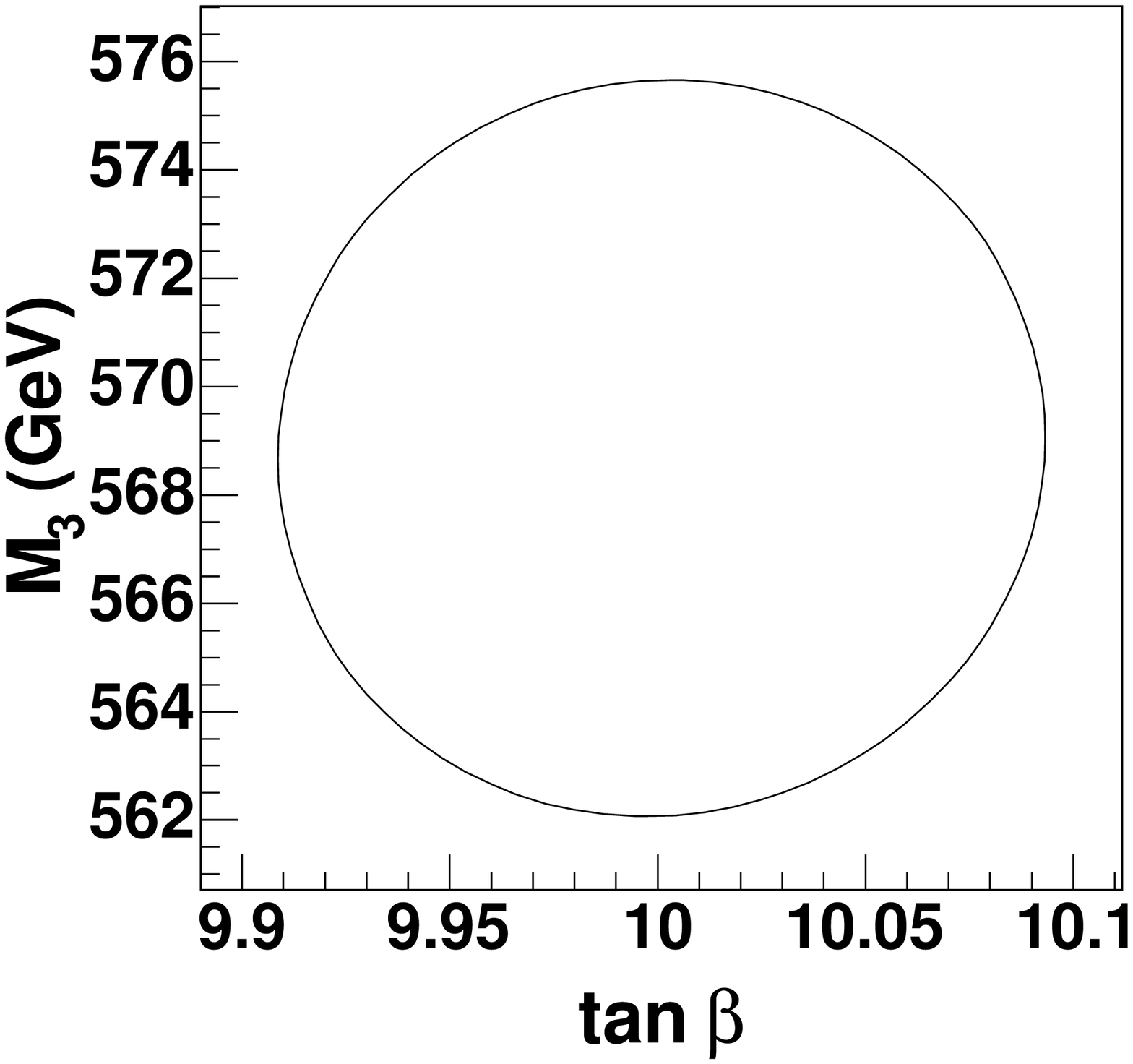}{0.49}
 \caption[Two-dimensional uncertainty contours of the Fittino SPS1a fit]{\sl Examples of 
   the two-dimensional uncertainty contours of the Fittino SPS1a' fit.}\label{fig:ellipse1}
\end{figure}

If the flag \texttt{UseMinos} is set, two sets of two-dimensional
uncertainty contours of all possible combinations of all parameters
are produced. In the first set, contained as two-dimensional
\texttt{TGraph} objects in the output file
\texttt{FitContours\_2params\_free.root}, all parameters but the two
plotted ones are kept fixed. In the second set, contained in
\texttt{FitContours.root}, all parameters are free, yielding larger
allowed areas inside the plotted $\Delta\chi^2=1$ area.
Fig.~\ref{fig:ellipse1} shows examples of the obtained curves for a
fit to the benchmark scenario SPS1a, which are ellipses in the case of
almost linear dependences of the observables on the parameters near
the fitted minimum. The slope of the ellipses indicates the parameter
correlations.

\subsection{Examples for the Pull Distributions}\label{sec:pull}

\begin{figure}[t]
  \placefigc{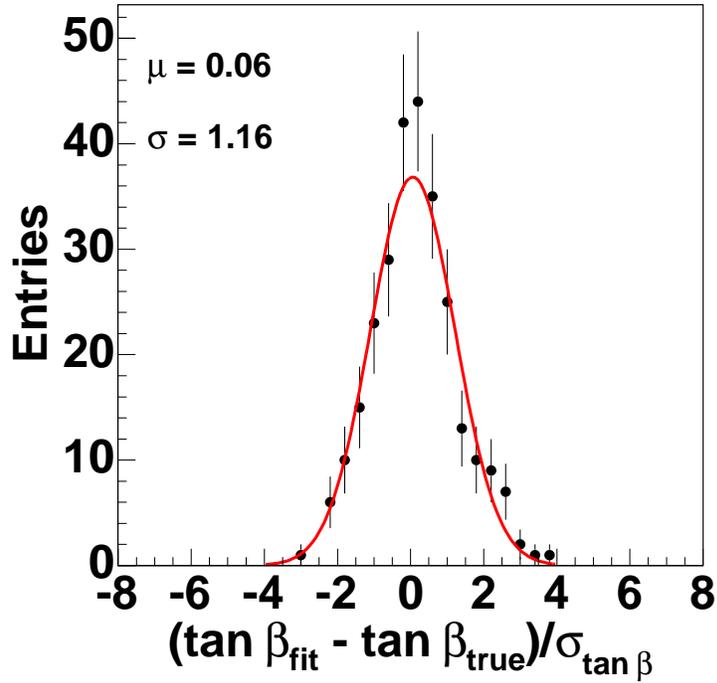}{0.65}
  \caption[Pull distributions of the parameters of the SPS1a fit, part I]{\sl Pull distributions 
    of the parameter $\tanb$ of the SPS1a' fit for 120 independent fits with
    observables smeared within their uncertainties. The uncertainties
    of all parameters are well described.  }\label{fig:pulldists1}
\end{figure}

\begin{figure}[t]
%  \placefigc{figures/chisq_distr_full_fit_040512.eps}{0.75}
  \placefigc{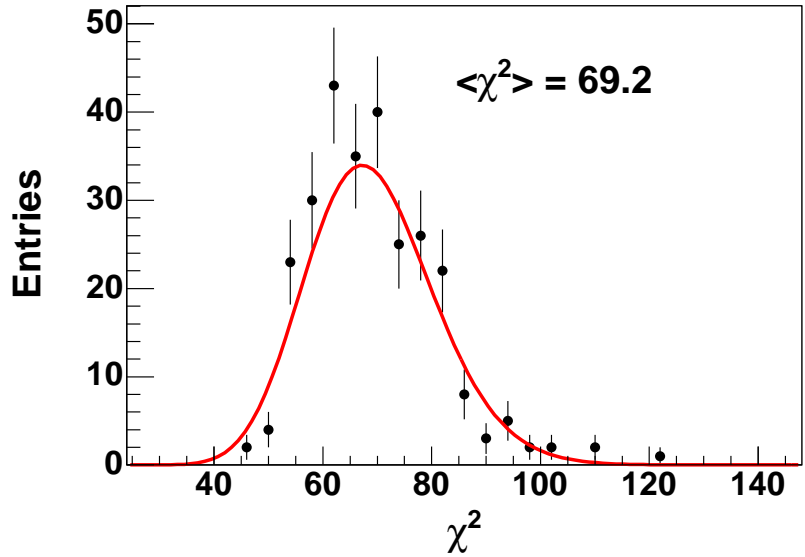}{0.75}
  \caption[Distribution of $\chi^2$ for the SPS1a' fit]{\sl Distribution of
    $\chi^2$ for the SPS1a' fit. 273 fits with 71 d.o.f.\ are
    performed.  The observables of the simulated measurements are
    smeared in each fit according to their uncertainties. The mean of
    the distribution is $\chi^2_{\mathrm{mean}}=69.2$, in
    agreement with the expectation of $71\pm0.9$.
  }\label{fig:chisq_distr_full_fit}
\end{figure}

If parameter values and uncertainties (as determined with Fittino) are
supplied for each parameter specified with \texttt{fitParameter}, pull
distributions and the $\chi^2$ distribution of the fit is
automatically calculated by Fittino if the flag \texttt{CalcPullDist}
is set. Examples for these distributions, as obtained in a fit in the
SPS1a scenario with 20 parameters, can be found in
Figures~\ref{fig:pulldists1} and \ref{fig:chisq_distr_full_fit}.

\clearpage

\subsection{Examples for the Observable Importance Determination}\label{sec:obs}

In the observable importance determination (using
\verb|CalcIndChisqContr|), the individual contribution of each
observable to the determination of each parameter is calculated.
Additionally, the total $\Delta\chi^2$ with respect to the minimal
$\chi^2$ for a variation of $\pm1\sigma$ of each parameter is shown.
If the total $\Delta\chi^2$ is large, then the parameter is strongly
correlated with other parameters. If it is close to 1, then there is
hardly any correlation. In the output file
\verb|fittino_individual_chisq_contr.out|, for each parameter first
all individual $\Delta\chi^2$ values from each observable are shown,
followed by the percentage of the contribution from the 5 largest
contributors. For the cross sections, the alias number is also shown
in order to ensure the identification of cross sections at different
polarisations and centre-of-mass energies.

\begin{verbatim}
 
Individual Delta Chisq contributions to Parameter TanBeta (10.0005 +- 0.333429)
================================================================

        list of all observables     :          Delta chi^2
                      .
                      .

                massNeutralino1     :             0.843007
                massNeutralino2     :              2.70032
                  massChargino1     :            0.0617443
                  massChargino2     :          0.000608651
ee  -> Neutralino1 Neutralino2 (19) :            0.0101081
ee  -> Neutralino2 Neutralino2 (20) :            0.0636789
 ee  -> SelectronL SelectronL~ (23) :           0.00277659
             ee  -> SmuL SmuL~ (25) :          0.000133514
           ee  -> Stau1 Stau1~ (27) :          0.000536534
           ee  -> Stau1 Stau1~ (28) :           0.00530975
    ee  -> Chargino1 Chargino1 (41) :            0.0489321
                   ee  -> Z h0 (45) :           0.00190326
                   ee  -> Z h0 (46) :            0.0012801
                      .
                      .

=========== percentage of strongest contributions: =============
                massNeutralino2     :             0.280837
ee  -> Neutralino2 Neutralino2 (20) :             0.191446
    ee  -> Chargino1 Chargino1 (41) :             0.126957
                massNeutralino1     :            0.0876738
ee  -> Neutralino1 Neutralino2 (19) :            0.0717215
              total Delta Chisq     :      9.615270

 
Individual Delta Chisq contributions to Parameter Mu (358.644 +- 1.14432)
================================================================
                      .
                      .
\end{verbatim}

%------------------------------------------------------------------------------------
%------------------------------------------------------------------------------------
% bibliography
%\clearpage
%\addcontentsline{toc}{section}{\numberline{}Bibliography}    
%\bibliography{fittino_literature}

\end{document}